\documentclass[a4paper,11pt]{article}
             
\usepackage{jheppub} 

\usepackage[T1]{fontenc} 
\usepackage{here}
\usepackage{grffile}
\usepackage{subfigure}
\usepackage{color}

\title{\boldmath Novel String Field Theory with also Negative Energy Constituents/Objects gives Veneziano Amplitude}

\author[a]{H. B. Nielsen}
\author[b]{and M. Ninomiya}

\affiliation[a]{Niels Bohr Institute, University of Copenhagen,\\17 Belgdamsvej, DK 2100 Denmark}
\affiliation[b]{Advanced Mathematical Institute,\\ Osaka City University, 3-3-138 Sugimoto, Sumiyashi-ku 558-8585, Japan}

\emailAdd{hbech@nbi.dk}
\emailAdd{msninomiya@gmail.com}

\abstract{We have proposed a new type of string field theory.
The main point of the present article is to cure some 
technical troubles:
missing two out three terms in Veneziano amplitude.
Our  novel 
string field theory, 
describes a theory with many strings in terms of ``objects'',
 which are not 
exactly, but close to Charles Thorn's string 
bits. The new point 
is that 
the objects in terms of which the universe states are 
constructed, and which have an essentially 26-momentum 
variable called $J^{\mu}$, can have the  
energy $J^0$ be also negative as well as positive. We get a long way 
in deriving in this model the Veneziano model and 
{\em obtain all the three terms } needed for a 
four point amplitude. This result strongly indicates
that our novel string field theory  {\em is} indeed 
string theory.}


\begin{document} 
\maketitle
\flushbottom

\section{Introduction}
In order to describe a situation with several strings\cite{1}, \cite{2}, \cite{3}, \cite{4}, \cite{5}, \cite{6}, \cite{7}
you need a priori a string field theory\cite{8} - second 
quantization - of the of the strings like in Kaku and Kikkawa\cite{9}, \cite{10}, \cite{11}, \cite{12} or
Witten theory\cite{13}, \cite{14}, \cite{15}, \cite{16}. We should cite the pioneering paper hew by Mandelstam\cite{17}, \cite{18}. Furthermore a seminal work on quantum string theory see\cite{19}. But we have ourselves rather 
a description\cite{21}, \cite{22}, Starting in advance more similar to the string bit
description of Charles Thorn et.al. \cite{23}, \cite{24}, \cite{25}, \cite{26}, \cite{27}, \cite{28}) in which the
state of an arbitrary number of strings is described 
by relating it to a state of a very large number of what 
we call ``objects'', and which have degrees of freedom like 
free particles. \\

The basic steps in writing in our formalism the second quantized
state/Hilbert vector for a given set of strings are the following:
\begin{enumerate}
\item
To every string construct the ``cyclic chain(s)''- one for an open string
and two for a closed one- in principle for each ``classical state'' by
wiriting the developing (single string) state in terms of the
splitting $X^{\mu}(\sigma,\tau)=X^{\mu}_{R}(\tau-\sigma)+X^{\mu}_{L}(\tau+\sigma)$.
Then the curves presented by $\dot{X}^{\mu}_{R}(\tau-\sigma)$ parametrized
by $\tau-\sigma$ and by $\dot{X}^{\mu}_{L}(\tau+\sigma)$ parametrized by
$\tau+\sigma$ in $25+1$ dimensional Minkowski space-time. 
For the open string there is a trick that actually these two curves
naturally continue each other into just \underline{one} closed curve/just
one closed cyclic chain.
\item
Next these ``cyclic chains'' (from 1.) are discretized into small
bits which we call objects. Notice that it is in the
``light cone'' variables $\tau_{R}=\tau-\sigma$ and 
$\tau_{L}=\tau+\sigma$ \underline{we} make discretization
into a lattice of ``objects'' (Not like Thorn theory which discrete  in $\sigma$).
Quantum mechanically some sophisticated trick is used only
the even numbered lattice points=objects (to avoid non-commutation
of the object variables).
\item
Next the many string state is represented by a Hilbert vector in an a priori
free massless scalar particle quantum field theory by acting on a certain
vacuum state $\mid 0 \rangle$ with a creation operator for each even
(numbered) ``object'' in any one of the set of strings the state of which
is to be constructed.\\
So the reader should see that we have made a correspondence which
to any state of an arbitrary number of strings let correspond a 
Hilbert vector in a massless free scalar quantum field
theory. I.e. we manage to make our novel string field
theory become an ordinary quantum field theory!
\end{enumerate}
\begin{figure}[H]  
\begin{center}
\subfigure[]{%
\includegraphics[clip, width=3cm]{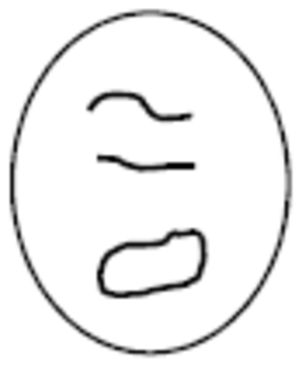}}%
\includegraphics[clip, width=1cm]{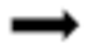} 
\subfigure[]{%
\includegraphics[clip, width=3cm]{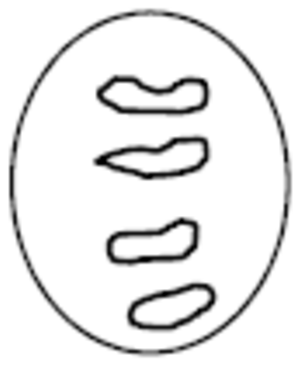}}%
\includegraphics[clip, width=1cm]{arrow.eps}
\subfigure[]{%
\includegraphics[clip, width=3cm]{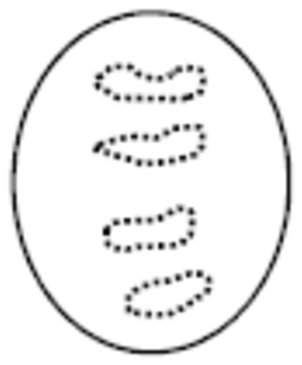}}%
\end{center}
\caption{Open and closed strings (a) to our object formalism (b) and (c)} 
\end{figure} 


From the previous works of ours it is important 
to remember:
We develop a formalism for description of an arbitrary 
number of strings (String Field Theory) by means of a 
Hilbert space formally with ``objects'' that are more 
like particles (they are ``bits'' in a technically a 
little different way from Thorn's theory. We explain the difference in sec.1.4.).\\
In terms of our ``objects'' (bits) the second quantized 
string world get totally static, scattering becomes a 
fake: the scattering amplitude becomes just the overlap 
of the initial with final states!\\
We get, after some technical procedure, the S-matrix = the overlap 
between initial and final string states to be the Veneziano 
amplitude\cite{29}.

It is very crucial that the objects must be able to have energy of \underline{both signs}.\\
So one piece of a cyclic chain can cancel another piece completely!\\
And thus pairs of compensating pieces of chains of objects may be phantazised where ever it may be.\\
``String'' comes ONLY in via the initial (and final state) conditions.\\
 
\subsection{Our SFT model Equivalent to String Theory} 
The final SFT model(string field theory model)\cite{20}, \cite{21}, \cite{22} of ours is described by the Fock space for massless 
non-interacting scalar bosons in $25+1$ dimensions.  That is to say it is described by a Hilbert space,
which is generated by a series of creation $a^{+}(\vec{p})$ and destruction $a(\vec{p})$
operators for scalar particels with 25-momenta $\vec{p}$, which can act successively on a
zero-particle state $\mid 0\rangle$.  That is to sey that the typical states in the Fock
space - or the Hilbert space describing the world state - are
\begin{equation}
a^{+}(\vec{p}_{1})a^{+}(\vec{p}_{2})\cdots a^{+}(\vec{p}_{n})\mid 0 \rangle
\end{equation}
\newpage
In the language, which we use, we call the scalar particles ''even objects'' and denote
their momenta by $J^{\mu}$ instead of $p^{\mu}$
(well really we only consider the conjugate momenta for the transverse components
$i=1,2,\ldots,24, i.e. J^{\mu}$).
\subsection{Relation To Charles Thorn's String Bits}
Our model / on string field theory is like they model for long considered prior to us by Charles Thorn also as a string bit theory in the sense that we discretize the strings. 
However, our way of discretizing deviates from the Thorn-version which makes the discretization string by discretizing the $\sigma$ -variable. But in oar case rather by discretizing separately right mover and the left mover string like systems. That is to say we First write string variable $X^\mu$ as a sum of a left mover and a right mover part such that $X^\mu(\sigma,\tau)=X^\mu_R(\tau-\sigma)+X^\mu_L(\tau+\sigma)$ and then AFTER THAT we perform the discretization by making what is essentially string bits but now for the variables $\sigma+\tau$ and $\sigma-\tau$ separately.  
 we then to distinguish can call ``objects'' -  
 instead after, we have split the solution into right and left mover and thus rather put into
  bits or now to distinguish objects  the right-mover variable ``$\tau$ -$\sigma$'' or the 
  left mover one ``$\tau+\sigma$''.\\
Actually Thorn has begun to do the same as we later at least for fermion modes.
\begin{figure}[H]
\begin{center}
\subfigure[For theory with only closed strings on the light cone.]{%
\includegraphics[clip, width=5cm]{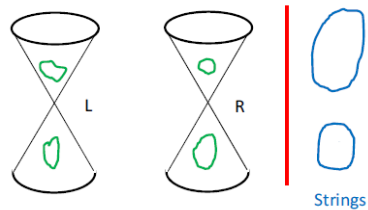} }%
\end{center}
\vspace{0cm}
\begin{center}
\subfigure[For theories with also open strings]{%
\includegraphics[clip, width=4cm]{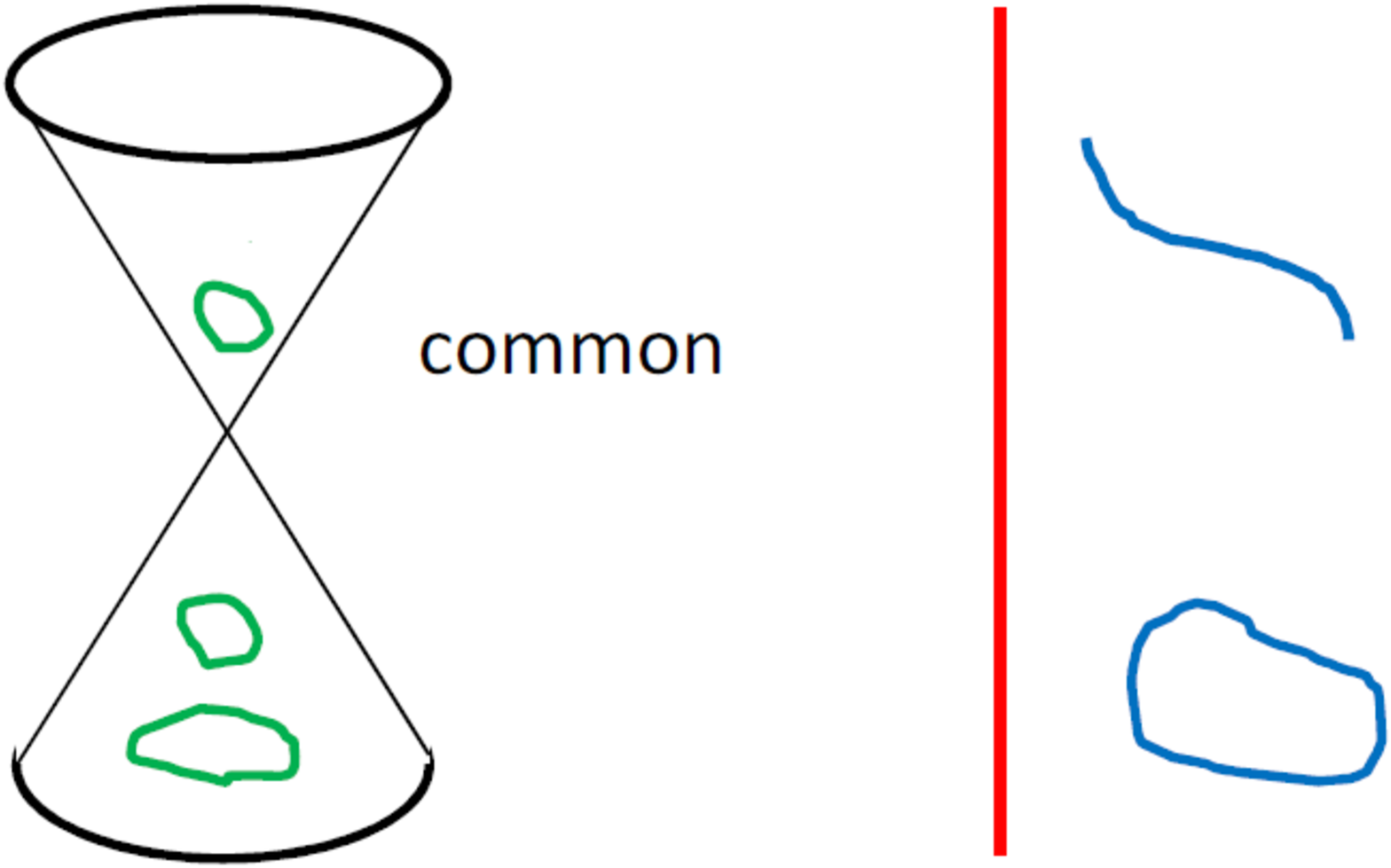} }%
\end{center}
\caption{Chains of objects on the light cone in the $25+1$ dimensional
Minkowski space. (a) is depicted on the closed strings
in terms of the left and right movers
(b) The left and right movers of the open and closed strings
are written together.} 
\end{figure} 
\newpage
\subsection{Translation from strings to ``Cyclic Chains''} 
\begin{figure}[H] 
\begin{center}
\subfigure[Open string corresponds to topologically circular figure: The cyclic chain]{%
\includegraphics[clip, width=5cm]{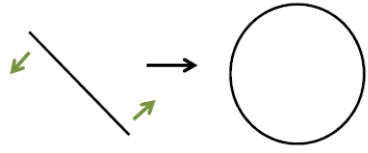} }%
\end{center}
\begin{center}
\subfigure[Closed string corresponds to 2 different cyclic chain]{%
\includegraphics[clip, width=5cm]{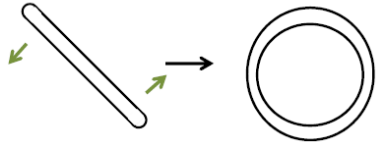} }%
\end{center}
\caption{Transition from strings to ``cyclic chains''} 
\end{figure} 


As a very simple example of a classically described open strings, we can think of a string rotating as a stiff stick around its middle point.  Say for simplicity that it is at rest and that we choose a gauge/a parametrization so that the energy is used to determine the parameter $\sigma$ along the string, so that an infinitesimal interval do in $\sigma$ just has the energy
\begin{equation}
dE=d\sigma
\end{equation}
If $r$ denotes the distance from the middle point $C$ to the point with coordinate $\sigma$, and $R$ denotes half the geometrical length (in target space) of the (open) string, and $\omega$ the rotation rate, then the energy density at the point $\sigma$ is
\begin{equation}
\frac{d\sigma}{dr}=\frac{dE}{dr}=\frac{1}{2\pi\alpha^{\prime}\sqrt{1-v^{2}}}
=\frac{1}{2\pi\alpha^{\prime}\sqrt{1-\omega^{2}r^{2}}} 
\end{equation}
Remembering
\begin{equation}
\frac{d}{d(\omega r)}arcsin(\omega r)=\frac{1}{\sqrt{1-(\omega r)^{2}}}
\end{equation}
thus
\begin{eqnarray}
d\sigma&=\frac{dr}{2\pi\alpha^{\prime}}\frac{d}{d(r\omega)}arcsin(\omega r)\\ \nonumber
&=\frac{1}{2\pi\alpha^{\prime}\omega}d\ arcsine(\omega r)
\end{eqnarray}
so that by integrating we may take the integration constant to get
\begin{equation}
\sigma=\frac{1}{2\pi\alpha^{\prime}\omega}arcsine(\omega r)
\end{equation}
or
\begin{equation}
r=\frac{1}{\omega}sin(\omega 2\pi\alpha^{\prime}\sigma)
\end{equation}
The half length $R$ of the (open) string is of course the maximal achievable
\begin{equation}
R=\frac{1}{\omega}.
\end{equation}
Let us identify the usual internal string time $\tau$ with the target space time 
essentially and let us take the plane in which the string rotates to be the 
$X^{1}X^{2}$-plane. Then while the coordinates
$X^{3}=X^{4}=\ldots=X^{25}=0$ we have
\begin{eqnarray}
X^{1}(\sigma, \tau)&=&cos(\omega 2\pi\alpha^{\prime}\tau)r\\ \nonumber
&=&\frac{1}{\omega}cos(\omega 2\pi\alpha^{\prime}\tau)sin(2\pi\alpha^{\prime}\omega\sigma)\\ \nonumber
X^{2}(\sigma, \tau)&=&\frac{sin(\omega 2\pi\alpha^{\prime}\tau)sin(2\pi\alpha^{\prime}\sigma\omega)}{\omega}
\end{eqnarray}
To have a hope of having made a proper conformal gauge choice we take $\tau=\frac{t}{2\pi\alpha^{\prime}}$.\\
If the solution ansatz which we have here made on physical ground is indeed a solution to the internal D'Alembertian equation
\begin{equation}
\left(\frac{\partial^{2}}{\partial\tau^{2}}-\frac{\partial^2}{\partial\sigma^{2}}\right)X^{\mu}(\sigma, \tau)=0
\end{equation}
it shall be possible to resolve the $X^{\mu}$ into right and left mover like
\begin{equation}
X^{\mu}(\sigma, \tau)=X^{\mu}_{R}(\tau-\sigma)+X^{\mu}_{L}(\tau+\sigma).
\end{equation}
Indeed it is easy to use the formulas for taking $sin$ and $cos$ for sums of variables such as 
\begin{equation}
sin\left(\omega\cdot 2\pi\alpha^{\prime}(\tau\pm \sigma)\right)=sin(2\omega\pi\alpha^{\prime}\tau)cos(2\omega\pi\alpha^{\prime}\sigma)
\pm cos(2\omega\pi\alpha^{\prime}\tau)sin(2\omega\pi\alpha^{\prime}\sigma)
\end{equation}
and
\begin{equation}
cos\left(\omega 2\pi\alpha^{\prime}(\tau+\sigma)\right)=cos(2\omega\pi\alpha^{\prime}\tau)cos(2\omega\pi\alpha^{\prime}\sigma)\mp 
sin(\omega 2\alpha^{\prime}\pi\tau)sin(\omega 2\alpha^{\prime}\pi\sigma)
\end{equation}
to rewrite e.g.
\begin{eqnarray}
X^{1}(\sigma, \tau)&=&\frac{1}{\omega}cos(\omega 2\pi\alpha^{\prime}\tau)sin(\omega 2\pi\alpha^{\prime}\sigma) \\ \nonumber
&=&\frac{1}{2\omega}\left(sin(\omega 2\pi\alpha^{\prime}(\tau+\sigma)\right)-sin\left(\omega 2\pi\alpha^{\prime}(\tau-\sigma)\right) \\ \nonumber
&=&X^{1}_{L}(\tau+\sigma)+X^{1}_{R}(\tau-\sigma)
\end{eqnarray}
where then
\begin{equation}
X^{1}_{R}(\tau-\sigma)=\frac{-1}{2\omega}sin\left(\omega 2\pi\alpha^{\prime}(\tau-\sigma)\right)
\end{equation}
and
\begin{equation}
X^{1}_{L}(\tau+\sigma)=\frac{1}{2\omega}sin\left(\omega 2\pi\alpha^{\prime}(\tau+\sigma)\right)
\end{equation}
Similarly we can write
\begin{eqnarray}
X^{2}(\sigma,\tau)&=&\frac{sin(\omega 2\pi\alpha^{\prime}\tau)sin(\omega2\pi\alpha^{\prime}\sigma)}{\omega}\\ \nonumber
&=&\frac{1}{2\omega}\left(sin(\omega 2\pi\alpha^{\prime}(\tau+\sigma)\right)+sin\left(\omega 2\pi\alpha^{\prime}(\tau-\sigma)\right)\\ \nonumber
&=&X^{2}_{L}(\tau+\sigma)+X^{2}_{R}(\tau-\sigma)
\end{eqnarray}
where
\begin{eqnarray}
X^{2}_{L}(\tau-\sigma)&=&\frac{1}{2\omega}sin\left(\omega 2\pi\alpha^{\prime}(\tau-\sigma)\right)\\ \nonumber
X^{2}_{R}(\tau+\sigma)&=&\frac{1}{2\omega}sin\left(\omega 2\pi\alpha^{\prime}(\tau+\sigma)\right)
\end{eqnarray}
At the ends of the string corresponding in the above notation to
$\omega 2\pi\alpha^{\prime}\sigma=\pm\frac{\pi}{2}$ we shall have that the tension in the string
\begin{equation}
``{\rm tension}"\propto X^{\prime \mu}(\sigma,\tau)=-\dot{X}_{R}(\tau-\sigma)+\dot{X}_{L}(\tau+\sigma)
\end{equation}
shall be zero, since there is nothing further out.\\
Calling
\begin{eqnarray}
\tau_{R}&=&\tau+\sigma\\ \nonumber
\tau_{L}&=&\tau-\sigma
\end{eqnarray}
We could write our above formulas
\begin{eqnarray}
X^{1}_{R}(\tau_{R})&=&-\frac{1}{2 \omega}sin(\omega 2\pi\alpha^{\prime}\tau_{R})\\ \nonumber
X^{1}_{L}(\tau_{L})&=&\frac{1}{2 \omega}sin(\omega 2\pi\alpha^{\prime}\tau_{L})\\ \nonumber
X^{2}_{R}(\tau_{R})&=&\frac{1}{2 \omega}sin(\omega 2\pi\alpha^{\prime}\tau_{R})\\ \nonumber
X^{2}_{L}(\tau_{L})&=&\frac{1}{2 \omega}sin(\omega 2\pi\alpha^{\prime}\tau_{L})
\end{eqnarray}
and boundary conditions of no tension becomes
\begin{equation}
\dot{X}^{\mu}_{L}\left(\tau_{L}=\tau+\frac{\pi}{2\cdot2\pi\alpha^{\prime}\omega}\right)\\ 
=\dot{X}^{\mu}_{R}\left(\tau_{R}=\tau-\frac{\pi}{2\cdot2\pi\alpha^{\prime}\omega}\right)
\end{equation}
the end of the string having $\sigma=-\frac{\pi}{2}$ obtain
\begin{equation}
\dot{X}^{\mu}_{L}\left(\tau_{L}=\tau-\frac{\pi}{2\omega 2\pi\alpha^{\prime}}\right)\\ 
=\dot{X}^{\mu}_{R}\left(\tau_{R}=\tau+\frac{\pi}{2\omega 2\pi\alpha^{\prime}}\right)
\end{equation}
For $\mu=1$ for example we see that these conditions are true because a shift
in the $\tau$-argument $\tau_{R}$ or $\tau_{L}$ by 2 times $\frac{\pi}{2\cdot\omega 2\pi\alpha^{\prime}}$
corresponds to a shift by $\pi$ in the argument of the sine and that gives just the sign shift needed
because $X^{1}_{R}(\tau_{R})=-\frac{1}{2 \omega}sin(\omega 2\pi\alpha^{\prime}\tau_{R})$
while  $X^{1}_{L}(\tau_{L})=\frac{1}{2 \omega}sin(\omega 2\pi\alpha^{\prime}\tau_{L})$.


Apart from some shift in the argument the ${X}^{\mu}_{R}$ and ${X}^{\mu}_{R}$ are basically
the same functional form - as we also see in our example - due to the boundary 
condition(s) at the end of the string.\\
Because we have at present a bit bad notation used
$\sigma=\frac{\pi}{2\cdot 2\pi\alpha^{\prime}\omega}$ at one of the ends we get in fact
\begin{equation}
\dot{X}^{\mu}_{L}(\tau_{L})=\dot{X}^\mu_{R}\left(\tau_{R}=\tau_{L}+\frac{\pi}{2\pi\alpha^{\prime}\omega}\right)
\end{equation}
If we had the end to have $\sigma=0$ we would have simply gotten
\begin{equation}
\dot{X}^{\mu}_{L}(\tau_{L})=\dot{X}^\mu_{R}\left(\tau_{R}=\tau_{L}\right) \ ({\rm for\ \sigma=0\ at\ end}).
\end{equation}
In any case we only need to use either $\dot{X}^{\mu}_{R}$ or $\dot{X}^{\mu}_{L}$
since they are of the same form.\\

\subsection{The closed string}

On the second translation on the figure we have taken as example a closed string in the configuration
wherein it is put back and forth along the same piece of line, and it - really meaning its two pieces -
rotate just like the open string just considered.  Now the seeming end points are just accidental but
not truly physically.  In this case of closed we shall \underline{not} identify the right and left movers.
Rather each of them give rise to its own ``cyclic chain''.  Therefore we have for this drawn
\underline{two} independent (although they happen to have the same coordinates) ``cyclic chains''.
Let us stress the rule: To an open string corresponds only \underline{one} cyclic chain, while to a 
closed string there corresponds \underline{two}, one for the right mover modes and one for the left.\\
Concerning the above discussion it should be noted that we considered
\begin{eqnarray}
\dot{X}^{\mu}_{R}(\tau_{R})&=&\frac{d X^{\mu}_{R}}{d \tau_{R}}\\ \nonumber
\dot{X}^\mu_{L}(\tau_{L})&=&\frac{d X^{\mu}_{L}}{d \tau_{L}}
\end{eqnarray}
rather than $X^{\mu}_{R}(\tau_{R})$ and $X^{\mu}_{L}(\tau_{L})$ themselves - and that is something we
for technical reasons, have decided to do in our Novel SFT - Had we not done that, the arguing away
having both $X^{\mu}_{L}$ and $X^{\mu}_{R}$ in the open string would \underline{not} have worked.
So it were a quite important point to make this differentiation!\\
%

For illustration of our formulation/model for our novel string field theory 
you shall imagine drawing in 25 or 26 dimensional perspective the  
right mover field $X^\mu_R$ differentiated with respect to its variable $\tau$ -$\sigma$, 
thinking classically.\\
To each open string right mover derivative is a 26-vector being a periodic function with the period 
used for $\sigma$.  The boundary condition at the end ensures that right and left mover derivatives 
are equal for the open string.\\
Thus we get to the open string a corresponding topologically circular figure the ``cyclic chain''.
For a closed string one can both imagine drawing the right and the left mover and they become 
two in general different closed curve images (= two ``cyclic chains''). 

For a single string in the ``conformal gauge'' we have the well-known equation of motion
\begin{equation}
\left(\partial^{2}_{\tau}-\partial^{2}_{\sigma}\right)X^{\mu}(\sigma,\tau)=0
\end{equation}
and solve it by the splitting
\begin{equation}
X^{\mu}(\sigma,\tau)=X^{\mu}_{R}(\tau-\sigma)+X^{\mu}_{L}(\tau+\sigma)=X^{\mu}_{R}(\tau_{R})+X^{\mu}_{L}(\tau_{L})
\end{equation}

As introduction to our Novel String Field Theory we shall imagine - and let us first think classically -
that for each string development in time - in \underline{Minkowski space} - draw a to such a moving/oscillating string
corresponding image of the $\tau_{R}=\tau-\sigma$ and $\tau_{L}=\tau+\sigma$
derivatives $\dot{X}^{\mu}_{R}(\tau_{R})$ and $\dot{X}^{\mu}_{L}(\tau_{L})$ of these 
 $X^{\mu}_{R}(\tau_{R})$ and $X^{\mu}_{L}(\tau_{L})$.\\
 Because of the periodicity for finite size strings the two images of $X_{R}$ and $X_{L}$ will be closed curves,
 called ``cyclic chains''.
 
\begin{figure}[H]
\begin{center}
\includegraphics[clip, width=7cm]{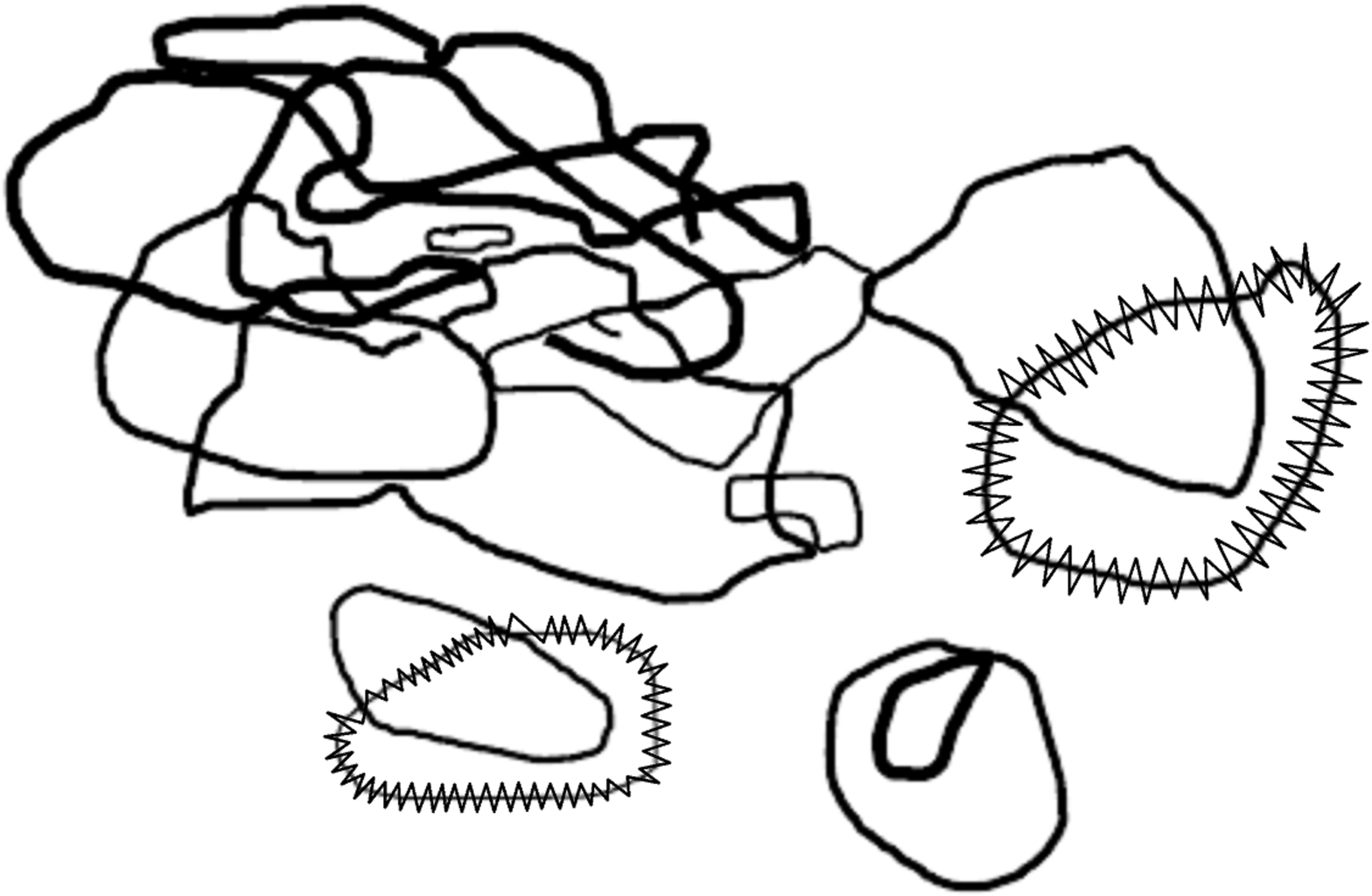}
\end{center}
\caption{The images of $X_{R}$ and $X_{L}$ are represented by the closed curves called cyclic chains. The all curves do not have ends.}
\end{figure}

The meaning of the foregoing figure with the net of curves is, that
there is in a perspectively drawn Minkowski space of $26=25+1$ dimensions (in bosonic string case).
A point on the net each time some one of all strings present in Universe for some value of
$\tau_{R}=\tau-\sigma$ or $\tau_{L}=\tau+\sigma$ the respective derivatives 
$\dot{X}^{\mu}_{R}(\tau_{R})$ or $\dot{X}^{\mu}_{L}(\tau_{L})$ of the string space
time position field $X^{\mu}(\tau,\sigma)$ take their vectorial value equal to that
point. This net of curves is thought classically at first: i.e. $\dot{X}^{\mu}_{R}(\tau_{R})$ and $\dot{X}^{\mu}_{L}(\tau_{L})$
have meaningful vectorial values once the last bit of gauge choice has been chosen. The reader is encouraged to first think of the net of curves ignoring quantum mechanisms, so that at least after a gauge choice the variables $\dot{X}^\mu_R$ and $\dot{X}^\mu_L$ have well defined values, that are $25+1$ vectors.

\subsection{One Open string would contribute say the slim closed curve contained in the net} 
\begin{figure}[H]
\begin{center}
\includegraphics[clip, width=7cm]{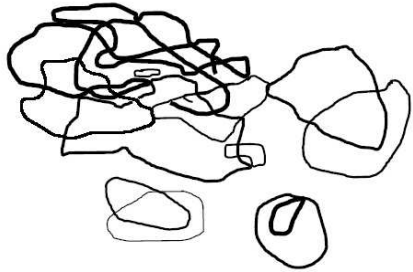}
\end{center}
\caption{One open string depicted by the slim closed curve contributes in the net.}
\end{figure}

For an open string one has at both end-points (say $\sigma=0,2\pi$)
\begin{equation}
X^{\prime\mu}=-\dot{X}^{\mu}_{R}+\dot{X}^{\mu}_{L}=0.
\end{equation}
Since this must be true for all $\tau$, it is enough information to deduce that for the open string
\begin{equation}
\dot{X}^{\mu}_{R}(\tau)=\dot{X}^{\mu}_{L}(\tau).
\end{equation}
Thus at any moment of $\tau$ there is for each open string a closed circle, a ``cyclic chain'', of 
$\dot{X}^{\mu}_{R}$ or $\dot{X}^{\mu}_{L}$, so that we get for each open string a closed circle
(the cyclic chain) of image points in the (perspectively imagined) $26=25+1$ dimensional space on the
figure.

\subsection{We Add one more - now wavy curve - Open  string Contribution, a Cyclic Chain} 
\begin{figure}[htbp]
\begin{center}
\includegraphics[clip, width=7cm]{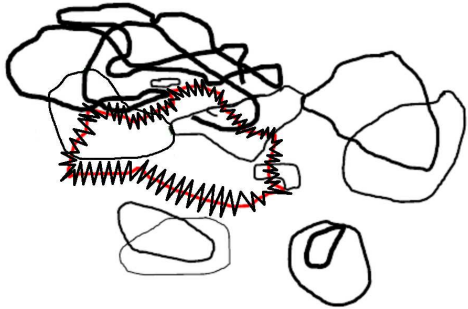}
\end{center}
\caption{One more open string contribution is added as a wavy closed curve in the cyclic chain. All curves have no ends.}
\end{figure}

\begin{figure}[htbp]
\begin{center}
\includegraphics[clip, width=7cm]{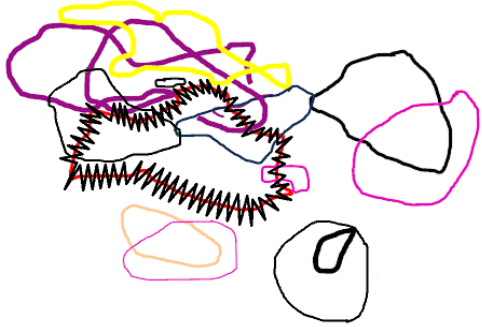}
\end{center}
\caption{The whole set of all the right or left mover derivatives are symbolized by the
thin-curved network. They can be interpreted as unification of a series of closed cycles that close each other.}
\end{figure}

\subsection{Comments on Our net of Cyclic Chains Representing String Via their Right and Left mover Derivatives} 
The whole set of all the right or left mover derivatives symbolized 
by the thin-curved network, can be interpreted as coming from various 
cyclic chains associated with various open string, or some pairs of 
them could correspond to closed strings.\\
We illustrated by various colors and thicknesses of the curves, 
how the image of all the right and left mover derivatives could be
divided into contributions e.g. from different open strings. 
(some pairs of cyclic chains potentially correspond to just the same 
closed string instead of to two open ones).\\
We stress that the division into contributions from different cyclic chains, 
different open strings say, is not unique !

\subsection{Further Comments on Novel String  Field Theory Model} 
For seeing our model, our novel string field theory, it is the crucial point to imagine that the cyclic chains shown on our figures here in the $25+1=26$ dimensional space(-time) should be thought of as series of ``objects'' meaning a discretization of the cyclic chains into particle-like objects, that can be created or annihilated by creation and annihilation operators.   So in this creation and annihilation we already now think quantum mechanically.\\
Thus the net  of cyclic chains is truly thought of as represented by the effect of a lot of actions 
with creation or annihilation operators on a ``background state''(=vacuum)\\   


The spirit of our novel string field theory is most quickly presented by
simply replacing the sites of $\dot{X}^{\mu}_{L}$ and $\dot{X}^{\mu}_{R}$
values representing a state of a universe with a number of open and
closed strings called Im by a representation by a Hilbert space vector
in a Hilbert space, in which what we call ``objects'' can be created
and annihilated. The main point is the replacement
\begin{equation}
Im \to \prod_{J^{\mu}_{\in}Im}a^{+}(J^{\mu})\mid ``vac"\rangle 
\end{equation}
written very shortly.\\
Here $J^{\mu}$ denotes a Minkowski space point and the product
runs over those $J^{\mu}$'s which lie in the union set Im.\\
There is, however, a series of technical troubles and the simple
replacement
\begin{equation}
Im \to \prod_{J_{\in}Im}a^{+}(J)\mid ``vac"\rangle
\end{equation}
is to be
considered an oversimplified pedagogical presentation, being 
quick correct.\\
First of all formulating a product of creation operators $a^{+}(J^{\mu})$
requires that the product runs over a discrete set rather than
a continuous set as $Im$. So really we rather should say
\begin{equation}
Im \to \prod_{I_{\in}{\rm discretized}\ Im}a^{+}(J^{\mu}(I))\mid ``vac"\rangle
\end{equation}
where we have split the ``continuous'' set $Im$ into a large
number of small pieces (or intervals in $\tau _{R}$ or $\tau_{L}$'s) enumerated by
essentially an integer $I$.\\
Secondly what we would achieve then would not be a true quantum theory
because $\dot{X}^{\mu}_{R}(\tau_{R})$ say in quantum single string theory does not
commute with itself for different $\tau_{R}$-values
\begin{equation}
[\dot{X}^{\mu}_{R}(\tau_{R}),\dot{X}^{\nu}_{R}(\tau^{\prime}_{R})]=i\delta^{\prime}(\tau_{R}-\tau^{\prime}_{R})
\end{equation}
To be able to properly quantize both the single object system and second quantize, the discretized 
$J^{\mu}(I)\stackrel{\propto}{\sim} \dot{X}^{\mu}_{R}(\tau_{R})$ should be so that they mutually \underline{commute}.
The trick to achieve such commutation since long proposed for our novel string field theory
was to only use to give creation operators $a^{+}(J^{\mu}(I))$ for the \underline{even} $I$'s.
That is to say that the true expression becomes
\begin{equation}
Im \to \left( \prod_{I_{\in}discretized\ Im\ and\ I\ even}a^{+}(J^{\mu}(I)) \right) \mid ``vac"\rangle 
\end{equation}
Here the important point is that the $I$ variable only runs over every other of the at first  
twice as large number of bits/objects. Because the commutator $i\delta^{\prime}(T_{R}-T^{\prime}_{R})$
after discretization gives lack of commutation between neighboring or objects we can achieve
full commutation if we leave out every other of the bits, we call the left out ones the odd and denote 
them by odd integers.\\
A third minor technical problem is that we should like to have the objects be their own
antiparticles (they should be ``Majorana'' so to speak) I.e.
\begin{equation}
a^{+}(J^{\mu})=a(-J^{\mu}).
\end{equation}
But if so we would with a ``normal'' quantization get
\begin{eqnarray}
a^{+}(J^{\mu}) \mid ``vac" \rangle=0\\ \nonumber
{\rm for}\ J^{0}<0.
\end{eqnarray}
In fact it were the trouble of our calculation 
 in which we only
obtained one of the three terms in the Veneziano amplitude. To avoid that we should even be able add 
some negative energy to the vacuum $\mid ``vac" \rangle$ i.e. we want
\begin{eqnarray}
a^{+}(J^{\mu})\mid ``vac" \rangle \neq 0\\ \nonumber
({\rm even\ for}\ J^{0}<0)
\end{eqnarray}
even when what we interpret  $J^{0}$ the energy is negative.\\
It is this problem that shall be solved by using as vacuum a state with the property that $a^{+}(J^{\mu}) \mid 0 \rangle$ shoud not be zero.
We call such a vacuum
``the rough Dirac sea''.


\subsection{The Rough Dirac Sea} 
At first one would be tempted to think of the vacuum or ``background state'' for  the objects as a state in which objects with 
positive energy  (if that makes any sense) were the only ones possible to produce, but...:\\
Even though energy of a single object can be given  a meaning, we shall assume that the ``background state'' for the second quantized objects-theory is not of the simple type that can only be modified to make the sum of the energies of the objects larger! Rather you can also add to it negative energy.\\
This is analogous to what we call the ``rough Dirac sea''(see Appendix)

On the last figure we show putting some arrows,  how to construct the points or small pieces of the string time track -i.e. the surface in Minkowski space through which the string in question passes: To each pair of small bits on the cyclical chain corresponds a little area on the space- time track of the string.\\
The string time track is a two-dimensional  manifold and thus one needs two one-dimensional parameters to parametrize it. We use for the open string the same  cyclic chain as being both parameters (two different point on the cyclic chain), while we for the closed string we use two different cyclic chains.

\subsection{Illustration of Connection to the String} 
We have put in on the picture fig.8 of the curve a narrow arrow. Such an arrow corresponds to a point on the string time track, or rather one point for each period of the string motion, in the sense that a couple of tangent vectors spanning the tangent to the string time track at the point in question are given by the two points in $25+1= 26$ space(-time) at the two ends of the (double)arrow.\\
To obtain all the time track points of an open string modulo periodicity you must take all the arrows that can connect two points on the cyclical chain describing the open string in question.\\
To obtain those for a closed string you must use all the arrows connecting one point on one of the two cyclic chains to the other one.

\begin{figure}[htbp]
\begin{center}
\includegraphics[clip, width=7cm]{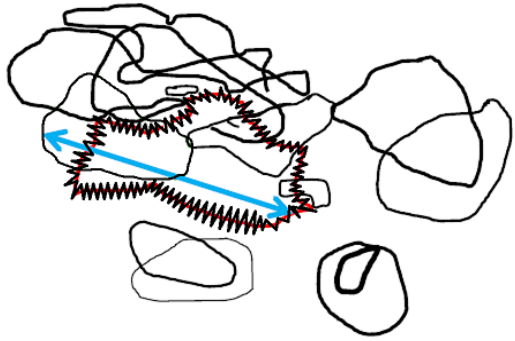}
\end{center}
\caption{Imagining putting the double around to corneot all rears of one point on the blue and one on the red it would come through all the time track points on a closed string. For each rear run through the $\dot{X}^{\mu}_{R}$'s on the two rounds flux the tangent plane to the time track of the closed string.}
\end{figure}

On the figure illustrating the closed string in terms of two cyclical chains you obtain the tangent- basis vectors of the various infinitesimal pieces of the closed string time track by going through all possible arrows connecting one point on the cyclic chain, green, and one on the cyclic chain, red.\\
Again we get a two-dimensional time-track of a string -now closed- by having it parametrized by two parameters running along cyclic chains.  \\

What happens if we first by creation or annihilation operator insert 
a piece of cyclic chain with one momentum distribution and then add
 the one with just the opposite one ? Actually they cancel and 
 it becomes as if nothing had been done.\\
This opens up a strange possibility for inserting the cyclic chains 
corresponding to a couple of say open strings: We could let a piece 
of the cyclic chain corresponding to one of the two strings happen 
to be just the ``opposite'' of a piece of the cyclic chain for 
the other open string.\\
In that case inserting the cyclic chains corresponding to the two open strings 
leads to there being  two pieces of cyclic chains canceling each other....\\
And thus the final state in our $25+1=26$ dimensional space for 
``objects'' would have got zero objects along the piece of cancellation. 
And the latter would not be marked in our Hilbert space for second quantized object..\\
One would only there ``see'' the pieces that were NOT canceled.

\begin{figure}[htbp]
\begin{center}
\includegraphics[clip, width=7cm]{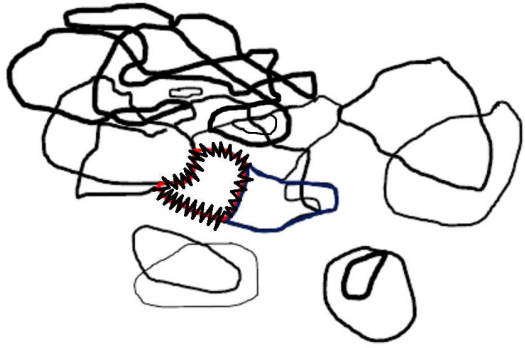}
\end{center}
\caption{Imagine that we move arround the blue double arrow in all possible positions connecting one point on the ``zigzagged'' loop and one point
on the ``thin'' loop. It then runs through a two-dimensional manifold
of pairs of end-points , and to each such pair of end-points (of the double arrow) the associated $\dot{X_R}^{\mu}$-pair specifies a twodimensional plane spanned by the pair of $\dot{X_R}^{\mu}$'s.
The set(manifold) of all these planes makes up the set of tangent planes
to the time-track (imbedded in the 25+1 dimensional physical
Minkowsky space-time) for the closed string development in question.}
\end{figure}

Having in mind the possibility of canceling two ``opposite pieces of cyclic chain we can without 
any trace in our second quantized object-state have two (oppositely oriented but otherwise 
locally in the same state) pieces of the blue and the red cyclic chains present.\\
On the above figure these canceling pieces are illustrated with the curve 
where the  blue and the red curves follow each other.\\
That there along this piece is no sign of the cyclic chains  in the object-description 
is illustrated by there being no thin  black curve along this piece before we drew 
in the red and the blue circles (cyclic chains) anywhere on top of 
the thin black curves illustrating the second quantized object-state.   


\section{The state of several strings}
At first -and that we did in our earlier work- in a Bled proceedings\cite{19}- 
one would simply take the second quantized Hilbert space state 
as the incomming two particle state $\mid\vec{p}_{1}, \vec{p}_{2}, inc \rangle$ 
to be the one obtained from the object formulation vacuum $\mid 0 \rangle$ 
by first acting with the creation operator for the cyclically ordered chain.
\begin{eqnarray}
C^{+}(string 1)=\int_{\begin{array}{cccc} around\\cyclic\\chain\\I_{e} even\end{array}}\Psi\left(J^{\mu}(0),J^{\mu}(2),\ldots,  J^{\mu}(N-2)\right) \\
\cdot a^{+}\left(J^{\mu}(0)\right)a^{+}\left(J^{\mu}(2)\right)\cdots a^{+}\left(J^{\mu}(N-2)\right)\prod _{I_{e}^{even}}{dJ(I_{e})}
\end{eqnarray}
on this vacuum $\mid 0 \rangle$ and then successively act with the analogous creation 
operator for string $2$, say $C$(string $2$). 
I.e. the description in our picture should be
\begin{equation}
C^{+}({\rm string} 2)C^{+}({\rm string} 1) \mid 0 \rangle
\end{equation}

\subsection{A formalism of Replacement of ghost}
If we imagine working with formulation with ghosts 
the $J^{\mu}(I_{e})$'s must be replaced by constructions 
such as $\left(J^{\mu}(I_{e}), \uparrow, \downarrow\right)$
or $\left(J^{\mu}(I_{e}), \downarrow, \downarrow\right)$ 
also involving ghost (but that is not so important just now) .\\
If one works with our old infinite momentum frame 
it would only be the transverse components 
and you would write instead
\begin{equation}
C^{+}({\rm string} 1)=\int\Psi\left(\vec{J}_{T}(0), 
\vec{J}_{T}(2),\ldots, \vec{J}_{T}(N-2)\right)
\prod_{\begin{array}{cccc}I_{e}\\even\\around\\circle \end{array}}
\left(a^{+}\vec{J}(I_{e})\right)\left(d\vec{J}_{T}(I_{e})\right)
\end{equation}

If our creation operators for the objects like 
$a^{+}\left(\vec{J}_{T}(I_{e})\right)$ all add say 
a positive energy rather than like the ones in say 
the BRST formalism $a^{+}\left(J^{\mu}(I_{e}), ghost\right)$, 
then there is no way that the products of a couple of creation 
operators for objects could be simplified.  
If we however have operators like 
$a^{+}\left(J^{\mu}(I_{e}), \uparrow, \downarrow\right)$ 
that are a priori able to bring energy and momentum of 
any sign, e.g. also negative energy $J^{0}$, 
then there is opened the possibility that the action of a couple of them
\begin{equation}
a^{+}\left(J^{\mu}(I_{e}), \cdot,\cdot\right)a^{+}\left(-J^{\mu}(I_{e}),\cdot,\cdot\right)\propto 1
\end{equation}
could act proportional to a c-number!\\
This is the crucial progress by not fixing energy nor longitudinal 
momenta to be positive. For this to work it is crucial that 
two opposite 26-J creation operators both are nontrivial. \\
i.e. both say
\begin{equation}
a^{+}(J^{\mu}, \cdot, \cdot) \mid 0 \rangle \neq 0
\end{equation}
and
\begin{equation}
a^{+}(-J^{\mu}, \cdot, \cdot) \mid 0 \rangle \neq 0
\end{equation}
That requires a ``rough Dirac sea''.
But provided - as we think we have - we have provided such nontrivial 
but opposite operator a ``Majorana'' boson theory for the objects say 
then one of the two operators $a^{+}(J^{\mu},\cdot, \cdot)$ and 
$a^{+}(-J^{\mu}, \cdot, \cdot)$ may be considered the annihilation 
operator for the particle / here object created by the other one.  
Thus indeed we may argue
\begin{equation}
a^{+}(-J^{\mu}, \cdot, \cdot)a^{+}(J^{\mu}, \cdot, \cdot) \mid 0 \rangle = \mid 0 \rangle
\end{equation}
so that the product acts as the unit operator.  
With continuum normalization one may have 
$\delta$-functions, but let us postpone this issue.\\
This means that in our ``rough Dirac sea'' 
picture we have to count that ``oppoiste''
(meaning opposite $J^{\mu}$, i.e. $J^{\mu}$ and $-J^{\mu}$ 
creation operators multiplied with each other can be replace 
-by calculation - by just unit operator.\\
\subsection{A trick of calculating the wave function}
Now we should remember that our crucial trick to calculate 
the wave functions $\Psi_{1}$, and $\Psi_{2}$ for our strings 
$1$and $2$ was to express them by means of imaginary $\tau $ 
functional integrals -so as to let only the ground state of 
the strings survive -.\\
If a couple of strings have it so that their associated 
creation operator cyclic chains can partly annihilate 
in the sense of giving unit operators as just described, 
then rather than being left in the two-string-describing 
state these objects can be removed provided they are ``opposite''.\\
The calculation of the amplitude for what the removing 
object creation operators can be well now by a functional 
integral for the complex (=imaginary $\tau $) developments 
(of the cyclically ordered chains) but with the extra rule: 
Piece(s) of the edge for string 1 half cylinder could be 
glued together with piece(s) of the edge for the 
half cylinder for string $2$:\\

\begin{figure}[H]
\begin{center}
\includegraphics[clip, width=6cm]{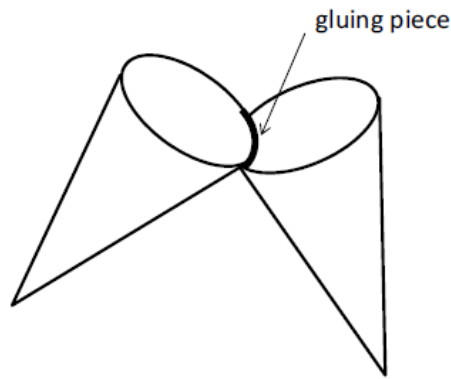}
\end{center}
\caption{The two cones can be glued together at the
common edges.}
\end{figure}

We would then -being closer to the classical 
solution - rather think of the picture\\

\begin{figure}[H]
\begin{center}
\includegraphics[clip, width=6cm]{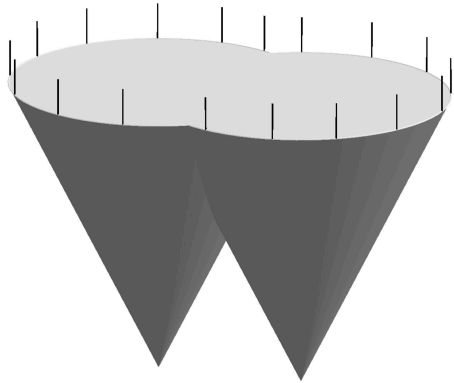}
\end{center}
\caption{The two cones can have a common edge as depicted.}
\end{figure}

as describing the functional integral to derive 
the resulting true object description for two string state
\begin{equation}
\mid {\rm str.}1, {\rm str.}2, \begin{array}{cc}in\\com\end{array}\rangle=C^{+}({\rm string}\ 2)C^{+}({\rm string}\ 1)\mid 0 \rangle
\end{equation}
The possibility that there were nothing annihilated 
should not be neglected, since it is not necessarily negligible.  
So rather symbolically we can write these two terms as on figure 12. \\

\begin{figure}[H]
\begin{center}
\includegraphics[clip, width=10cm]{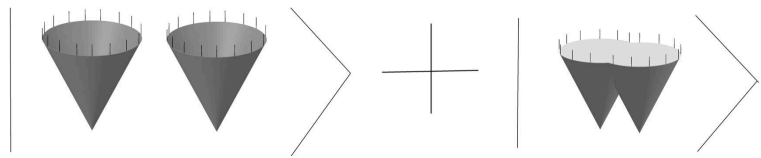} 
\end{center}
\caption{Two string states are sum of two state vectors as expressed in the figure.}
\end{figure}

These more complicated terms are obtained by taking from 
each of two incoming strings; string $1$ and string $2$ 
more than one piece of their cyclic chains and then we use the rule
\begin{equation}
a^{+}(J^{\mu}, \cdot, \cdot)a^{+}(-J^{\mu}, \cdot, \cdot)\sim 1 \label{J,.,.}
\end{equation}
along these -more than one- pieces.  
The typical construction such a more complicated term 
-namely corresponding to two pieces along which 
the rule (\ref{J,.,.}) 
 is used. It would lead to a contribution to 
the Hilbert space for the second quantized even object 
given by a functional integral for $J$'s (or$\Pi$'s) being 
fields on a two dimensional Euclidean manifold looking like\\

\begin{figure}[H]
\begin{center}
\includegraphics[clip, width=10cm]{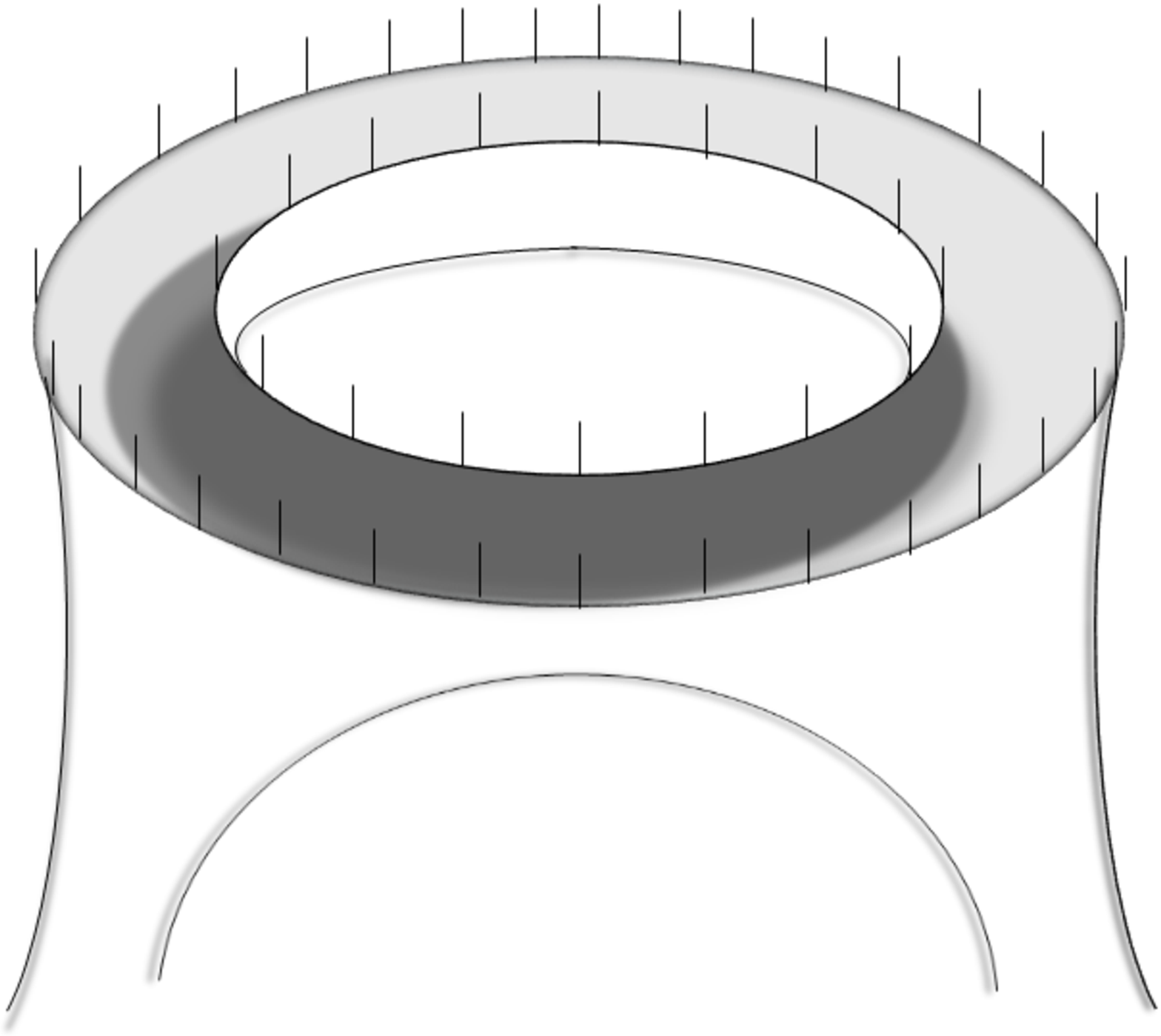} 
\end{center}
\caption{Topologically different one obtained by making use of
two cones in the manner described in main text.}
\label{fig.13}
\end{figure}

This drawing is meant to be obtainable topologically 
from the one for piece on which (\ref{J,.,.}) 
were used 
only by taking a two dimensional piece of the ``bottom'' 
pull it up while it hangs together and then chop off 
the cap of the pull out.  The piece of Hilbert vector\\

\begin{figure}[H]
\begin{center}
\includegraphics[clip, width=7cm]{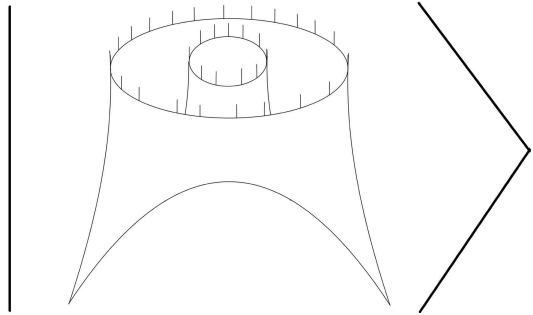} 
\end{center}
\caption{The Hilbert vector corresponding Figure \ref{fig.13}, to be included when 
calculating S-matrix element.}
\end{figure}
The two cyclic chainshas now to be identified with 
corresponding ones in the final state if we shall use our 
$\mid {\rm string} 1,  {\rm string} 2, inc.\rangle$
to make an $S$-matrix element.
%
\section{An Encouraging Step;The Veneziano model except for the integration weight}
Having argued for functional -integral constructions 
as suggested by the above and with imaginary time, 
we have really obtained, what is usually considered 
the correct description of string development in a 
double description.  The functional integrals are 
in our present philosophy just a mathematical trick 
to describe the wave function for the string states, 
while if you consider the functional integrals 
the usual way the surfaces drawn rather represent 
the development of the strings, 
even if in some way Wick rotated though.  One shall 
note that in our philosophy there is in our 
object formulation no development.  It is only 
a trick to construct the wave function.  
Nevertheless, there is an obvious similarity 
between the diagrams describing functional integrals 
which \underline{we} obtain when we write down or 
describe the overlap between an initial $\mid i \rangle$ 
and a final $\mid f \rangle$ states in terms of the 
functional integrals and to the same $S$-matrix element 
corresponding development diagrams.\\
From this actually quite perfect correspondence 
we can in fact as a beginning derive that we must 
obtain the Veneziano\cite{29} model(s) in question up to an extra factor under the integration sign. 
In the integral expression for the Veneziano model 
is more delicate to obtain.
%
We believe that this inclusion of such pieces of cyclic chains cancelling 
each other so that no track of their existence is left in our description 
by means of objects can be of help to solve a problem, which we met in our 
development of our string field theory based on such object description:\\
The problem were the following: We sought to calculate from our novel string 
field theory the scattering of two particles into two others expecting to obtain 
a Veneziano model with three terms corresponding to the usual three pairs among 
the channeles s, t, and u.  But ${\bf \cdots}$\\
We got only one of these terms!

We did the calculation, that turned out in this way unsuccessfully 
in an infinite momentum frame ``gauge''(=parametrization) choosing 
the right mover and the left mover coordinates ensuring a fixed amount 
of the ``longitudinal'' component of the $26$-momentum for all objects.\\
Thus there were in this ``gauge'' choice no way to have the 
``longitudinal'' component of momentum made opposite.\\
So there were no way in that ``gauge'' to realize the 
``Phantasy or cancellation pieces of cyclic chain''.\\
All cyclic chains corresponding to (open) strings would have to be 
``visible'' in the second quantized object description.\\
But then two strings cannot become one by partial annihilation in the cyclic chain description .   

In principle our ``Novel String Field Theory'' should just be a 
rewriting of a system of many strings interacting
with each other.  There should be nothing logically new, only reformulation!  Whether we really have logical perfect 
correspondence (after quantization) depends, however, on how much information we count it that there is in our
formalism.  Strictly speaking we could make ``philosophically different versions'' of our model, each including
different amounts of information in them.  Only the one with large amount of information would match usual string
theory.  But we suggest to take the version with minimal amount of information most serious as our novel string field theory.

\section{The Classical Approximation Summary of ``Layers of Existence Degrees''} 
But in our formulation, the ``Novel String field Theory'' 
it is we think pedagogical to consider
several {\it layers of truth or existence} corresponding 
to different versions with respect to including
information into the formalism.  Let us first describe 
these existence -  ``layers'' in our classical
(by classical formulation we have in mind that the single particle states are described classically
with both $\Pi$ and $J$ having values simultaneously - in disagreement with Heisenberg uncertainty
principle - while the objects are still second-{\it quantized}, so that a state in superposition of
having different numbers of objects is in the picture) formalism, with which we started.

We have a series of steps from truly existing in our novel theory to being more and more phantasy, not 
really existing:
\begin{enumerate}
\item
Fully existing the system of objects that can have both negative and positive energy - because they sit 
on a background of the ``rough Dirac sea'' (which is also fully existent, although we avoid having
to go in detail formulating it.)
\item
Chaining of Objects into Cyclical chains from the continuity of the strings and the boundary
conditions we have the objects forming cyclically ordered chains with objects sitting with
neighbor distances of the order of the ``latticification cut off distance''.  Really we do not
take it in our model that this chaining order has any physical existence in itself; but seeing
a pattern of the ``truly'' existing objects with their $J^{\mu}$'s and $\Pi^{\mu}$'s we may let the nearness
define for us a chaining. (A wrong way of chaining may lead to bad continuity)
\end{enumerate}

\begin{enumerate}
\item
Fully existing
\item
Chaining of Objects into Cycles
\item
Pairing by this ``pairing'' we mean the information telling, which cyclically
ordered chains together corresponds to a string.  Open strings come from just one
cyclically ordered chain each, while closed strings each need two, but here the total 
26-momenta for the two shall be the same.  So again knowing the cyclic chains there
is some basis/restriction for guessing, which ones to combine.
\item
Cancelling pieces the phantasy pieces of opposite $26$-momenta just invented.
\end{enumerate}

As told: We first attempted a description with the single object -and also single string- 
being treated classically, but allowing quantum mechanics in the second quantization, 
so that we could make a superposition of even different numbers of objects or strings.\\
While in Thorn's bits from pieces of sigma 
 have positions commuting with each other, 
the right-mover part of the position does NOT commute with itself. 
Rather there is for its derivative -which we want to work with a delta-prime function commutator.\\

If we shall have seperate creation and annihilation operators objects in any state, it would at least be a very unwanted complication if the degrees of 
freedom for one object and another one did not commute.  Thinking classically on the single object state we should thus have {\bf zero Poisson 
bracket between the variables associated with two different objects.} This is, however, {\bf impossible} if we want the objects to represent at least 
the $\tau$-derivative of say the right mover part of the string position field $\dot{X}^{\mu}_{R}(\tau_{R})$, because these rightmover
fields or their derivatives for different values of the argument $\tau_{R}$ ($\tau_{R}$ is usually replaced 
by a complex variable $\bar{z}$ ) do {\bf not commute}, thus do not have zero Poisson bracket.\\
 
Wanting
\begin{equation}
J^{\mu}_{R}(\tau_{R}) \propto \dot{X}^{\mu}_{R}, 
\end{equation}
we define
\begin{equation} 
J^{\mu}_{R}({\rm for \ interval \ bit} [\tau_{R}-\Delta/2,\tau_{R}+\Delta/2])=X_{R}(\tau_{R}-\Delta/2)-X_{R}(\tau_{R}+\Delta/2),
\end{equation}
where $\Delta$ is our cut off ``length'' in ``bits'' or ``objects''.\\
But these $\dot{X}^{\mu}_{R}$'s do not commute, but rather
\begin{equation}
\left[\dot{X}^{\mu}_{R}(\tau_{R}),\dot{X}^{\mu}_{R}(\tau^{\prime}_{R})\right]=i\delta^{\prime}(\tau^{\prime}_{R}-\tau)\neq 0
\end{equation}

\begin{figure}[htbp]
\begin{center}
\includegraphics[clip, width=7cm]{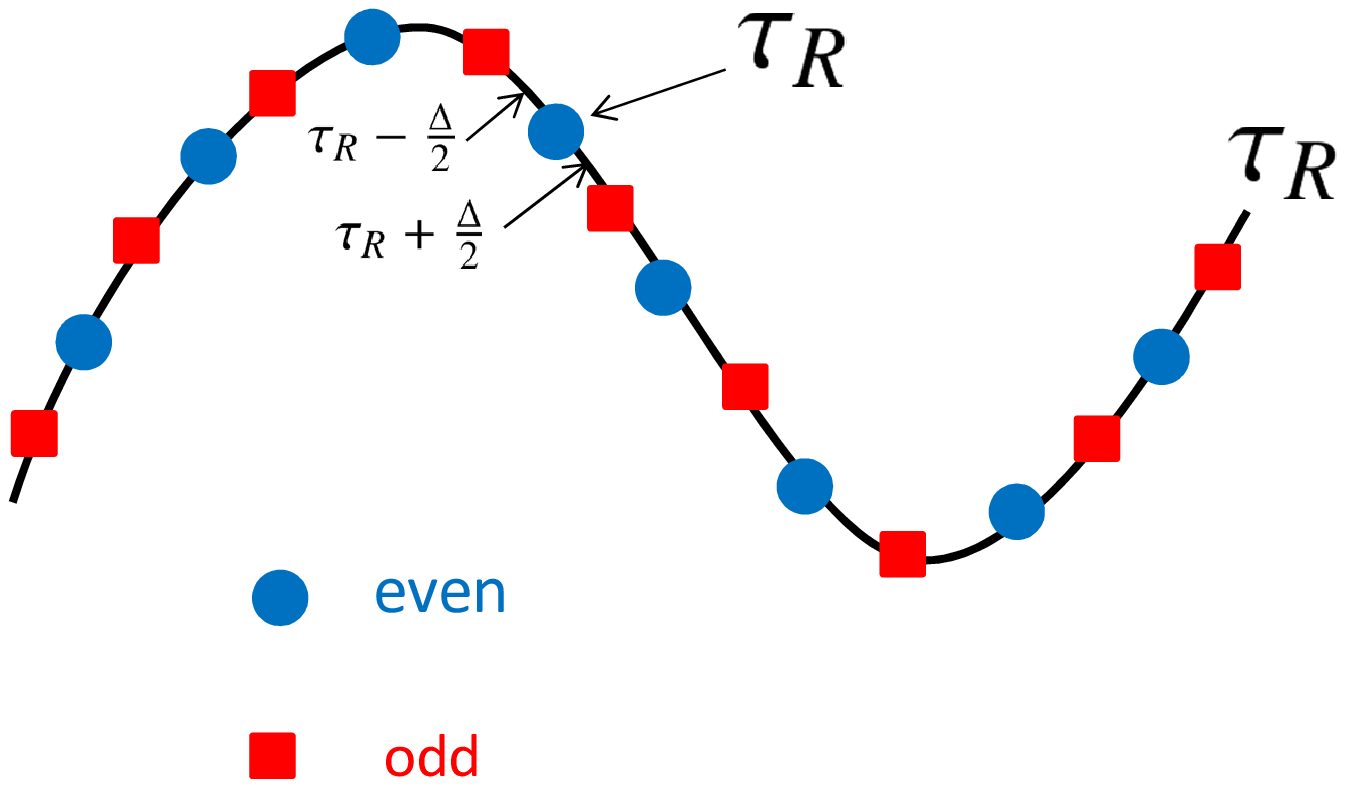}
\end{center}
\caption{The string consists of a series of even and odd objects along the string
alternatively put. As is depicted as an example the string is parametrized
as the Figure 15.
The right mover is parametrized by $\tau_{R}$. If the one even 
object is in $\tau_{R}$, the neighboring two objects are odd ones:
one of them is situated at $\tau_{R}-\frac{\Delta}{2}$ and the other is
at $\tau_{R}+\frac{\Delta}{2}$, where $\Delta$ is the cut off length
into objects.}
\end{figure}

Replacing the $\tau_{R}$ parametrization of right movers - replacing $\bar{z}$ which were complex - by a discrete counting $I$ taking
{\it integer} values our discretized object variables
\begin{equation}
J^{\mu}_{R}(I)=X^{\mu}_{R}\left(\tau_{R}(I)+\Delta/2\right)-X^{\mu}_{R}\left(\tau_{R}(I)-\Delta/2\right),
\end{equation}
where say $\tau_{R}(I)={\rm constant}+I\ast a$, obey crudely at least
that even $I$ object variables $J^{\mu}_{R}(I)$ commute with each other, and
the odd ones commute with themselves, 
However the even ones do NOT commute with their TWO ODD neighbors!\\
This is just describing a discretized delta-prime function.\\

We dueled in constricting our SFT to only consider the even numbered objects as independent objects. Then we  let the variables - especially $J_{R}$ for the
ODD objects be written in terms of the conjugate variables $\Pi^\mu(I)$ of the neighboring even object variables:
\begin{equation}
J^{\mu}_{R}(I)({\rm for\ odd}\ I)=-\alpha^{\prime}\pi\left(\Pi^{\mu}(I+1)-\Pi^{\mu}(I-1)\right).
\end{equation}
the $\Pi(I\pm1)$ to be used here are numbered by the even numbers $I\pm1$ 
and thus can be the conjugate momenta of the assigned $J^{\mu}_{R}(I\pm1)$ 
to the even objects respectively. 

We can consider this expression for the odd $J_{R}$ as a kind of 
``integrating up'' the odd $J_{R}$ to construct/give us the $\Pi$'s.
For an even value of the integer $K$ we solve our prosed equation for the odd $J_{R}$'s
\begin{equation}
\Pi^{\mu}_{R}(K)=\frac{J^{\mu}_{R}(K-1)+J^{\mu}_{R}(K-3)+J^{\mu}_{R}(K-5)+\cdots}{-\alpha^{\prime}\pi}.
\end{equation}
It looks that with this ``integrating up'' information on the original continuum string variable
$\dot{X}^{\mu}_{R}(\tau_{R})\sim J^{\mu}_{R}$ has been moved away in a non-local way for odd
discrete points and is stored as the $\Pi^{\mu}_{R}$ for even argument.\\
It may be less serious though since $\dot{X}^{\mu}_{R}(\tau_{R})\sim J^{\mu}_{R}$ were already
differentiated, so really the $\Pi^{\mu}_{R}$ becomes essentially the right mover part of the
position variable for the string.\\

If our so called string field theory is only a theory of essentially 
free massless objects, then it is a mystery: where is the string?\\
Do we even get the Veneziano model out of it? Yes we do. 
We actually can calculate to obtain Veneziano model -actually though 
for an OVERLAP between initial and final state rather than for a complicated S-matrix.

\section{Main idea in Calculating Veneziano Amplitude} 
\begin{enumerate}
\item
Since nothing goes on the S-matrix can only be unity and the S-matrix element just an overlap of in and outgoing states $\langle f \mid i\rangle$.
\item
We write these in or outgoing states by having for each particle a wave function in terms of objects.
\item
These wave functions are written by means of an IMAGINARY TIME functional integral for a STRING extracting the ground state (of the string) by it surviving  long imaginary time development.
\end{enumerate}


We wrote of the overlap becoming the scattering aplitude as the
various pieces of surfaces with corresponding functional 
integrals, that were used to deliver the wave functions in
object formulation corresponding to the incomming string, 
and to the outgoing ones. By looking at the various cases
we see that for the
\begin{equation}
1+2\to 3+4
\end{equation}
scattering of ground-state (tachyons, in bosonic string theory), we
realize at first that the combined surfaces modulo
conformal and topology conserving transformations consists
of an $S^{2}$-sphere with the four external ground state
particles/string being attached to this $S^{2}$-sphere;

\begin{figure}[H]
\begin{center}
\includegraphics[clip,width=10cm]{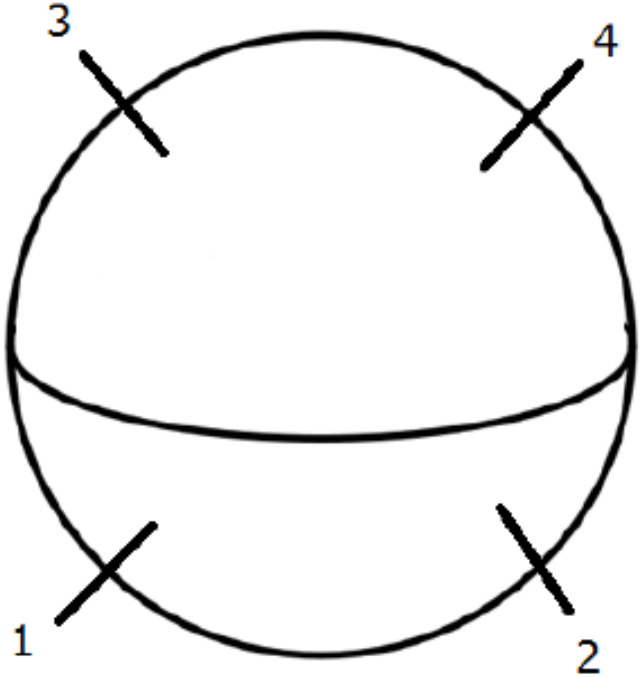} 
\end{center}
\caption{Illustration of four ground state scattering
$1+2\to3+4$ where the surface is an $S^{2}$-sphere.}
\label{S2-sphere}
\end{figure}

Because of the conformal invariance -up to an anomaly- of the funcitonal 
integral the possible inequivalent configurations of the four
external lines on the Riemann-sphere $S^{2}$ is given by a
single complex anharmonic ratio
\begin{equation}
A=\frac{z_{1}-z_{3}}{z_{2}-z_{3}}:\frac{z_{1}-z_{4}}{z_{2}-z_{4}}
\end{equation}
where the $z_{i}$'s are the Riemann surface notation places for
the four external lines. By considering the construction in 
more detail one can see that in fact this anharmonic ratio $A$
becomes real and that all possible real values of $A$ can become
realized. By the three values
\begin{equation}
A=0,1,\infty 
\end{equation}
corresponding to that a couple of external line attachments
coincide the real axis compactified to the topology of an 
$S$ circle is divided into three intervals:

\begin{figure}[H]
\begin{center}
\includegraphics[clip,width=5cm]{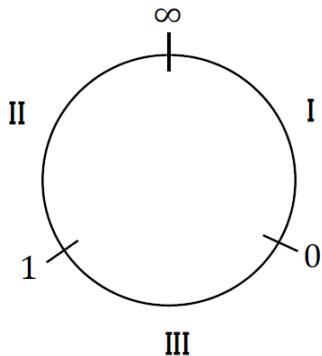} 
\end{center}
\caption{Inequivalent configurations of the four
external lines on the Rieman-Sphere $S^{2}$ are
$A=0,1,\infty$. The external line attachments is 
divided.  $S^{2}$ topology's circle is divided into
three intervals: I, I\hspace{-1pt}I and I\hspace{-1pt}I\hspace{-1pt}I.}
\end{figure}

Each of these three pieces would, provided we obtain the right 
external- momentum- independent- measure $\rho$ say, give
rise to a term in the scattering amplitude as there are in the
Veneziano model
\begin{equation}
g^{2}\Bigl(B\bigl(-a(t),-\alpha(s)\bigr)+B\bigl(-\alpha(s),-\alpha(u)\bigr)
+B\bigl(-\alpha(u),-\alpha(t)\bigr)\Bigr).
\end{equation}
For instance with an appropriate definition of an anharmonic ratio of
the type of $A$, namely
\begin{equation}
X_{\mu t}=\frac{z_{1}-z_{4}}{z_{1}-z_{2}}:\frac{z_{3}-z_{4}}{z_{3}-z_{2}}
\end{equation}
where we use the enumeration of external lines for $1+2\to 3+4$ and
call the $14-$ or $23-$ channel the $t$-channel
\begin{equation}
(-p_{1}+p_{4})^{2}=(-p_{2}+p_{3})^{2}=t
\end{equation}
and the $13-$ or $24-$channel the $u$-channel
\begin{equation}
(-p_{1}+p_{3})^{2}=(-p_{2}+p_{2})^{2}=u
\end{equation}
we obtained in our previous article \cite{22} that we got such
an extra measure $\rho$ that the amplitude piece coming from
\begin{equation}
0\le X_{\mu t}\le 1
\end{equation}
because
\begin{equation}
\int^{1}_{0}X_{\mu t}^{-\alpha_{t}(t)^{-1}}(1-X_{\mu})^{-X_{u}(u)^{-1}}
\rho(X_{\mu t})d\alpha_{\mu t}=B\left(-\alpha(t),-\alpha(u)\right).
\end{equation}
Our hope is that, an evaluation of the external momentum
independent factor $\rho$ for the two other regions in the 
anharmonic ratio, will turn out to deliver the two missing terms,
$B\left(-\alpha(s),-\alpha(t)\right)$ and $B\left(-\alpha(s),-\alpha(u)\right)$.

We should stress that we have already by referring to usual string scattering theory
that the factor in the integrand depending on the external momenta
has the correct form for giving these missing terms. This is already extremely
promissing for that we shall obtain the full Veneziano model.

Let us also stress that in fact the two missing terms come from those terms in the 
overlap which in the object representation have parts of cyclic chains corresponding
to different, say incoming strings just \underline{compensating}/\underline{annihilating}
each other. So the two missing terms in our previous paper only come about,
because we now have -due to rough Dirac sea- (See Appendix A and B) the possibility of cyclic chain pieces
with also negative energy.\\

\subsection{The simplest case of only positive $J^{+}=\frac{a \alpha^{\prime}}{2}$ 
(even) objects:}
Here as a starting example of calculating Veneziano model we consider
the case that two incoming \underline{open} strings have only \underline{postive}
$J^{+}=\frac{a \alpha^{\prime}}{2}$ objects in their cyclic chains, and 
also the two outgoing strings $3$ and $4$ have only positive $J^{+}$ objects.\\
Then all the objects of the cyclic chain of string $1$ must go to either $3$
or $4$ and the separation with fewest breakpoints (points where
one class of objects stop and a new class begins) means that the objects on
the cyclic chain of string $1$ just falls into two pieces. One of these two pieces go into the final state string 3, the other one goes into string 4.\\


\begin{figure}[H]
\begin{center}
\includegraphics[clip, width=5cm]{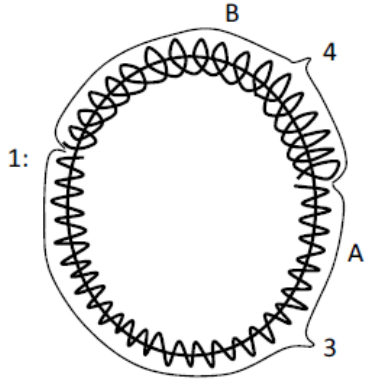}
\end{center}
\caption{All the objects of the cyclic chain of
string $1$ must go to either $3$ or $4$ and the
separation objects on the cyclic chain of string $1$
fall into two pieces $3$ or $4$.}
\end{figure}

By saying the same for string $2$ you soon end up with scheme fig. 19.\\

\begin{figure}[H]
\begin{center}
\includegraphics[clip, width=7cm]{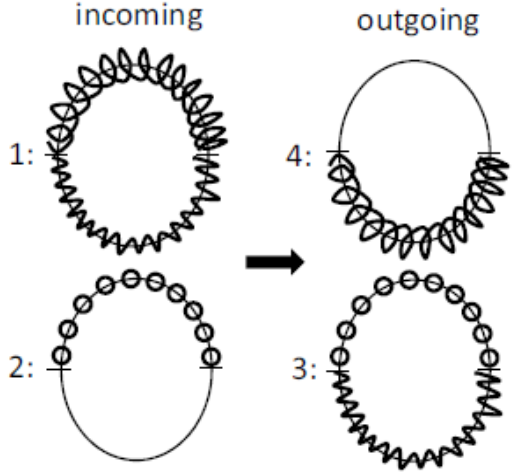}
\end{center}
\caption{These figures illustrates how we think of the four cyclically ordered chains corresponding to the two incoming, 1 and 2, and the two outgoing, 3 and 4, strings in the scattering process, are divided into pieces according to where they are going or coming from. That is to say we denoted by one signature that series of objects on the cyclically ordered chain corresponding to string 1 which go into the cyclically ordered chain corresponding to the outgoing string 4, say. Similarly the other pieces going to or coming from a specific cyclically ordered chain and going to a specific other one have got a distinguishable signature. The arrow symbolizes the transition from initial to final states. }
\end{figure}

How many objects there are in the four different classes 
marked by their curve signatures 
\includegraphics[clip,width=1cm]{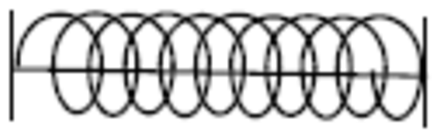},
\includegraphics[clip,width=1cm]{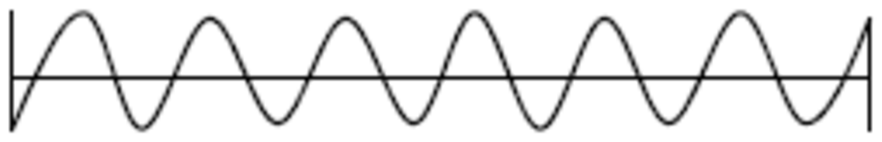},
\includegraphics[clip,width=1cm]{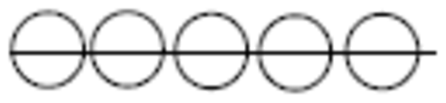}, and
\includegraphics[clip,width=1cm]{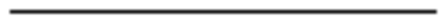}
are denoted by 
$N_{\includegraphics[clip,width=1cm]{helix.eps}}$,
$N_{\includegraphics[clip,width=1cm]{wavy.line.eps}}$,
$N_{\includegraphics[clip,width=1cm]{ring.eps}}$,
$N_{\includegraphics[clip,width=1cm]{line.eps}}$
respectively and we have assuming the easily derivable rule
that the 26-momentum
of the open string
\begin{equation}
P^{\mu}_{string}=\frac{1}{2\pi\alpha^{\prime}}\sum^{N-1}_{I=0}J^{\mu}(I)
\end{equation}
that the ``longitudinal momenta'' of the four strings (We use a light cone or metric notation with $\eta^+=\eta^0+\eta^{25}$ and $\eta^-=\eta^0-\eta^{25}$)
\begin{eqnarray}
P^{+}_{string\ 1}=\frac{a\alpha^{\prime}}{2\Pi\alpha^{\prime}\cdot 2}
\left(N_{\includegraphics[clip,width=1cm]{helix.eps}}
+N_{\includegraphics[clip,width=1cm]{wavy.line.eps}}\right)
&=&\frac{a}{4}\left(N_{\includegraphics[clip,width=1cm]{helix.eps}}
+N_{\includegraphics[clip,width=1cm]{wavy.line.eps}}\right)\\ \nonumber
P^{+}_{string\ 2}&=&\frac{a}{4}\left(N_{\includegraphics[clip,width=1cm]{ring.eps}}
+N_{\includegraphics[clip,width=1cm]{line.eps}}\right)\\ \nonumber
(*)P^{+}_{string\ 3}&=&\frac{a}{4}\left(N_{\includegraphics[clip,width=1cm]{ring.eps}}
+N_{\includegraphics[clip,width=1cm]{wavy.line.eps}}\right)\\ \nonumber
P^{+}_{string\ 4}&=&\frac{a}{4}\left(N_{\includegraphics[clip,width=1cm]{line.eps}}
+N_{\includegraphics[clip,width=1cm]{helix.eps}}\right)
\end{eqnarray}
Having in mind that we shall calculate the Veneziano amplitude for one set of 
external 26-momenta at a time, we must get the amplitude for one such fixed
set of external momenta with a sum running over of the various possible combinations
($N_{\includegraphics[clip,width=1cm]{helix.eps}}$,
$N_{\includegraphics[clip,width=1cm]{wavy.line.eps}}$,
$N_{\includegraphics[clip,width=1cm]{ring.eps}}$,
$N_{\includegraphics[clip,width=1cm]{line.eps}}$)
obeying these equations (*) for the fixed external momenta.
This means obviously that the variations under the summation obey
\begin{equation}
\Delta N_{\includegraphics[clip,width=1cm]{helix.eps}}=
-\Delta N_{\includegraphics[clip,width=1cm]{wavy.line.eps}}=
+\Delta N_{\includegraphics[clip,width=1cm]{ring.eps}}=
-\Delta N_{\includegraphics[clip,width=1cm]{line.eps}}.
\end{equation}
Thinking of the 
$N_{\includegraphics[clip,width=1cm]{helix.eps}}$,
$N_{\includegraphics[clip,width=1cm]{wavy.line.eps}}$,
etc. as true numbers of objects, they cannot become negative.
One should have in mind that there is only one free variable say 
$N_{\includegraphics[clip,width=1cm]{helix.eps}}$ to sum over.
To achieve the most simple range over which to sum we would much
like that 
$N_{\includegraphics[clip,width=1cm]{helix.eps}}$ and
$N_{\includegraphics[clip,width=1cm]{ring.eps}}$ 
would run down to $0$ simultaneously and that
$N_{\includegraphics[clip,width=1cm]{wavy.line.eps}}$ and
$N_{\includegraphics[clip,width=1cm]{line.eps}}$
would also go to zero simultaneously, in the
opposite end of the integration/summation so to speak.
But from (*) this wish would imply
$P^{+}_{1}=P^{+}_{3}$ and $P^{+}_{2}=P^{+}_{4}$ (from
$N_{\includegraphics[clip,width=1cm]{helix.eps}}=N_{\includegraphics[clip,width=1cm]{ring.eps}}=0$
simultaneously), while 
$N_{\includegraphics[clip,width=1cm]{wavy.line.eps}}=N_{\includegraphics[clip,width=1cm]{line.eps}}=0$
simultaneous imply $P^{+}_{1}=P^{+}_{4}$ and $P_{2}^+=P^+_{3}$.
All together these wish-requirements thus mean
\begin{equation}
P^{+}_{1}=P^{+}_{2}=P^{+}_{3}=P^{+}_{4} \label{wish-requirements}
\end{equation}
If we trusted that the theory of ours were Lorentz invariant
we might achieve this equality (\ref{wish-requirements}) by an appropriate 
Lorentz transformation.\\
If we do not have this case of equal $P^{+}$'s there can be
different ways of connecting the objects from initial to final
state and it may get more complicated.\\
Then under the assumption of the ``longitudinal momenta''
$P^{+}_{\kappa}$ for $\kappa=1,2,3,4$ being all equal
we obtain just a summation over say
$N_{\includegraphics[clip, width=1cm]{helix.eps}}$ 
running over all integers from $0$ to
$4a^{-1}P^{+}_{1}=4a^{-1}P^{+}_{2}=\ldots=4a^{-1}P^{+}_{4}$. \\
We shall approximate this summation by an integration
\begin{equation}
\sum_{I\in [0,4a^{-1}P^{+}_{1}]}\to \int dN_{\includegraphics[clip, width=1cm]{helix.eps}}
\end{equation}
and from here obtain the integration so characteristic for the Euler's Beta function
form of cntributions to the Veneziano model.\\
A lot of details of the set up of the calculational
procedure is to be found in \cite{5}, \cite{6}, \cite{7}.
The main point is that one first write the 
wave function for the set of even objects
-$N_{\includegraphics[clip, width=1cm]{wavy.line.eps}}+N_{\includegraphics[clip, width=1cm]{wavy.line.eps}}=4P_{string}/a$
of them, or rather only $2P_{string}/a$ of them when 
we only want the even ones- by being given by the
value of a functional integral being 
a regularized (cut off e.g. by a lattice)
Feynman-Dirac-Wentzel one with the Nambu action
or better one already made into a conformal 
gauge form
\begin{equation}
\int exp\Bigl(-(2\pi\alpha^{\prime})^{-1}\int_{A}\bigl(\vec{\partial}\phi^{\mu}
(\sigma^{1},\sigma^{2})\bigr)d\sigma^{1}d\sigma^{2}\Bigr)D\phi^{\mu}. \label{conformal_gauge_form}
\end{equation}
(The index $\mu$ is the Lorentz index; but basically one develop
the different factor with different $\mu$ separately; at the end multiply
then.)\\
$A$ is chosen to be either a (unit) disk with the center punctured out or what 
is via conformal transformations a half infinite cylinder.\\
-By this conformal transformation the punctured center
corresponds to the cylinder running to infinity end-\\

\begin{figure}[H]  
\begin{center}
\subfigure[A unit disk with the center punctured out]{%
\includegraphics[clip, width=5cm]{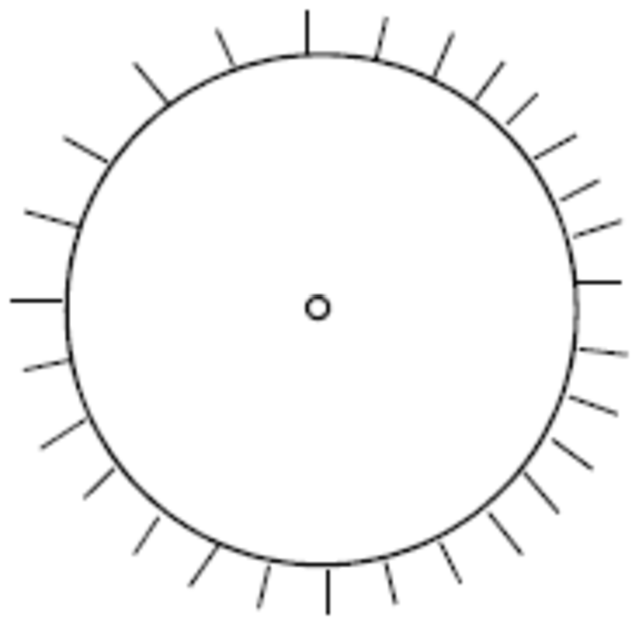}}%
\hspace{1cm}
\subfigure[A half infinite cylinder]{%
\includegraphics[clip, width=5cm]{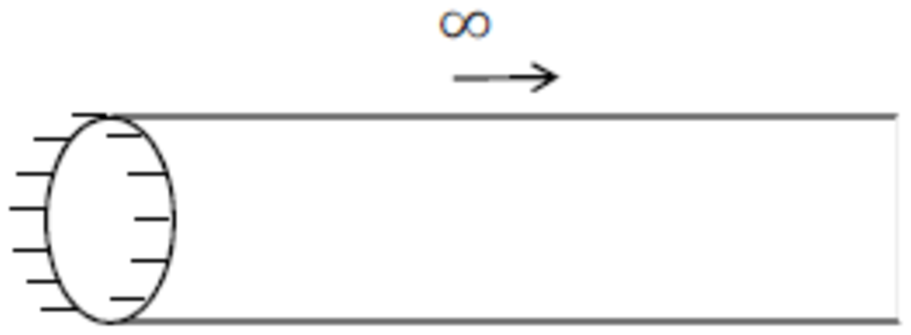}}%
\end{center}
\caption{(a) and (b) are conformally equivalent} 
\end{figure}

Along the edge of the disk or conformally
equivalent the finite end of the half infinite cylinder
the objects in the cyclic chain corresponding to 
one of the strings, say string $1$, are imagined attached in a
equidistant.\\
The meaning of this attachment to the edge is that the values
$J^{\mu}(I)$ associated with objects around the cyclic chain
are identified with differencec of neighboring values of
the functional integral (dummy) variable $\phi^{\mu}$ 
(see equation (\ref{conformal_gauge_form}). If there are in total $N$ objects
in the cyclic chain the attached object $I$ may be at the
angle -along the disk-edge-
$\theta(I)=2\pi\cdot\frac{I}{N}$ and then we put (say)
\begin{equation}
J^{\mu}_{R}(I)\stackrel{=}{\rm ident.}
\phi ^{\mu}\left({\rm exp}\left(i2\pi\frac{I+1}{N}\right)\right)
-\phi^{\mu}\left({\rm exp}\left(i2\pi\frac{I-1}{N}\right)\right) 
\end{equation}
for all even $I$. Because of our technique of only taking the
even objects as truly existing and being directly used in the
second quantized description in our Hilbert space we only
use the even $I$ values here.\\
Here the complex numbers exp $i2\pi\frac{I\pm 1}{N} $
refers to that the disk-shaper region A for the
functional integral is imbedded into the complex plane 
as the unit disc (and so the norm unity numbers lie on
the edge of the disc.)\\
The reader can relatively easily see first that the 
functional integral comes to depend on the object
$J^{\mu}$ variables in a ``Gaussian'' (exponential of a
quadratic expression in these $J^{\mu}(I)^{\prime}s$) way
and then secondly that the ``fluctuations'' in the various
Founrier resolution coefficients of these $J^{\mu}(I)^{\prime}s$
as a function of the angle $2\pi\frac{I}{N}$ become the same
as these fluctuations as estimated e.g. in our article
on the mass spectrum of the string in our
object-formulation \cite{22}.\\
But the easiest way to see that the functional 
integral gives the wave function is by thinking of producing the ground state by propagating  
during an imaginary time span. If this time span is taken to be long, 
go to infinity, the dominantly surviving string state will be the ground state.\\
This is illustrated above 
 by the figure\\

\begin{figure}[H]
\begin{center}
\includegraphics[clip, width=5cm]{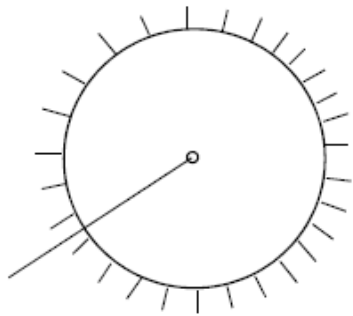}
\end{center}
\caption{The ground state of the string state is 
obtained in the functional integral by propagating
during an imaginary time span. If this time span
is taken to be long, e.g. go to infinity, the 
dominantly surviving string state is the ground state.}
\end{figure}

Now we have four external particles giving rise to four such
wave functions which can be written by means of discs or half-finite region
functional integrals.\\
We shall mark the objects sitting along the edges of 
these discs by the symbols on fig.21
.\\
Having in mind that we can as well as a disc use the complement of a disc depicted as Figure 22.\\
\begin{figure}[H]
\begin{center}
\includegraphics[clip, width=8cm]{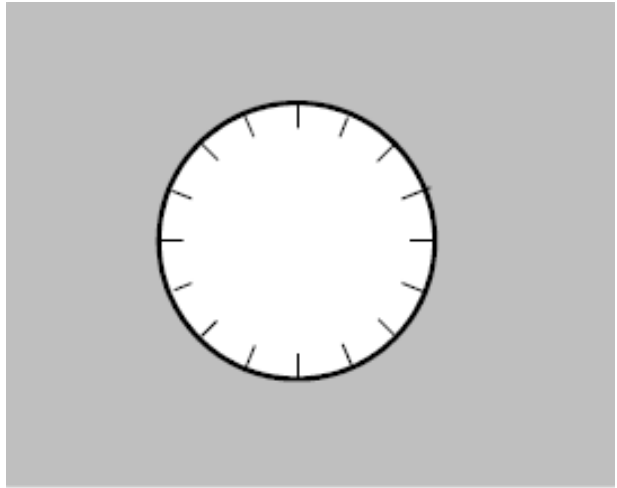}
\end{center}
\caption{The complement of disc. Four external particles that give rise to
four wave functions. They are written by means of
discs or half-finite region functional integrals.
The objects sit along the edges of the discs. We can use
as well as disc's complement of a disc. In fact the
region outside the disc is the two dimensional region. On this figure we have drawn two layers with the four discs put in as discs and complements.}
\end{figure}
\newpage
The meaning of the region is that outside a disc as the 2-dimensional region
we can figure out a way 
%
to match the regions of objects marked with corresponding
symbols, even in such a way that all four discs or complements lay on two
layers in the complex plane/Riemann sphere.
In fact we shall put them so that 1 and 2, i.e. the incoming particles/strings
have their wave functions represented by (proper) discs -lying
in the two layers respectively -while the outgoing strings
3 and 4 are represnted by complements of also
unit discs lying correspondingly by in the
two layers over the complex plane.\\
In this way we get the whole complex plane covered doubly,
one representative in each layer all over.
\begin{figure}[H]
\begin{center}
\includegraphics[clip, width=15cm]{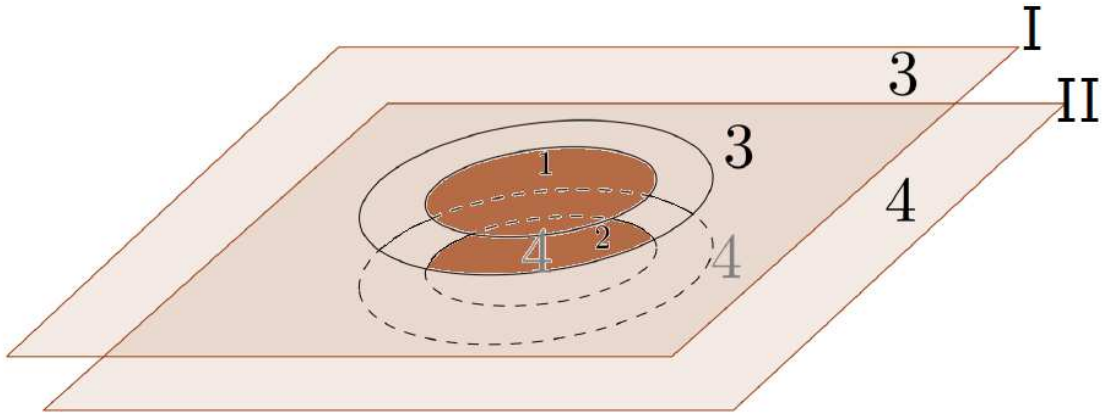}
\end{center}
\caption{The two lines structure.}
Fig.Caption: Layer first one is denoted I while the second is II. These two layers are connected.
\end{figure}
To be definite we can choose to put in the first layer, 
called I, the disc for string $1$ and the complement of the 
disc for string $3$.\\
We put them so that the common edge for these
two regions \includegraphics[clip, width=1cm]{wavy.line.eps}
are just placed side by side (on the unit circle). 
Since by our assumption of all four strings having
the same longitudinal momenta
$P^{+}_{K}$ so that 
$P^{+}_{1}=P^{+}_{2}=P^{+}_{3}=P^{+}_{4}$.
We easily saw that 
$N_{\includegraphics[clip, width=1cm]{wavy.line.eps}}=N_{\includegraphics[clip, width=1cm]{line.eps}}$,
and so we can analogous in the second layer, called I\hspace{-1pt}I, put the pieces
marked with \includegraphics[clip, width=1cm]{line.eps}
in the edges for 2 and 4 just covering the same part of
the unit circle (just now on layer I\hspace{-1pt}I) as the piece marked
connecting 1 and 3.

\begin{figure}[H]
\begin{center}
\includegraphics[clip, width=15cm]{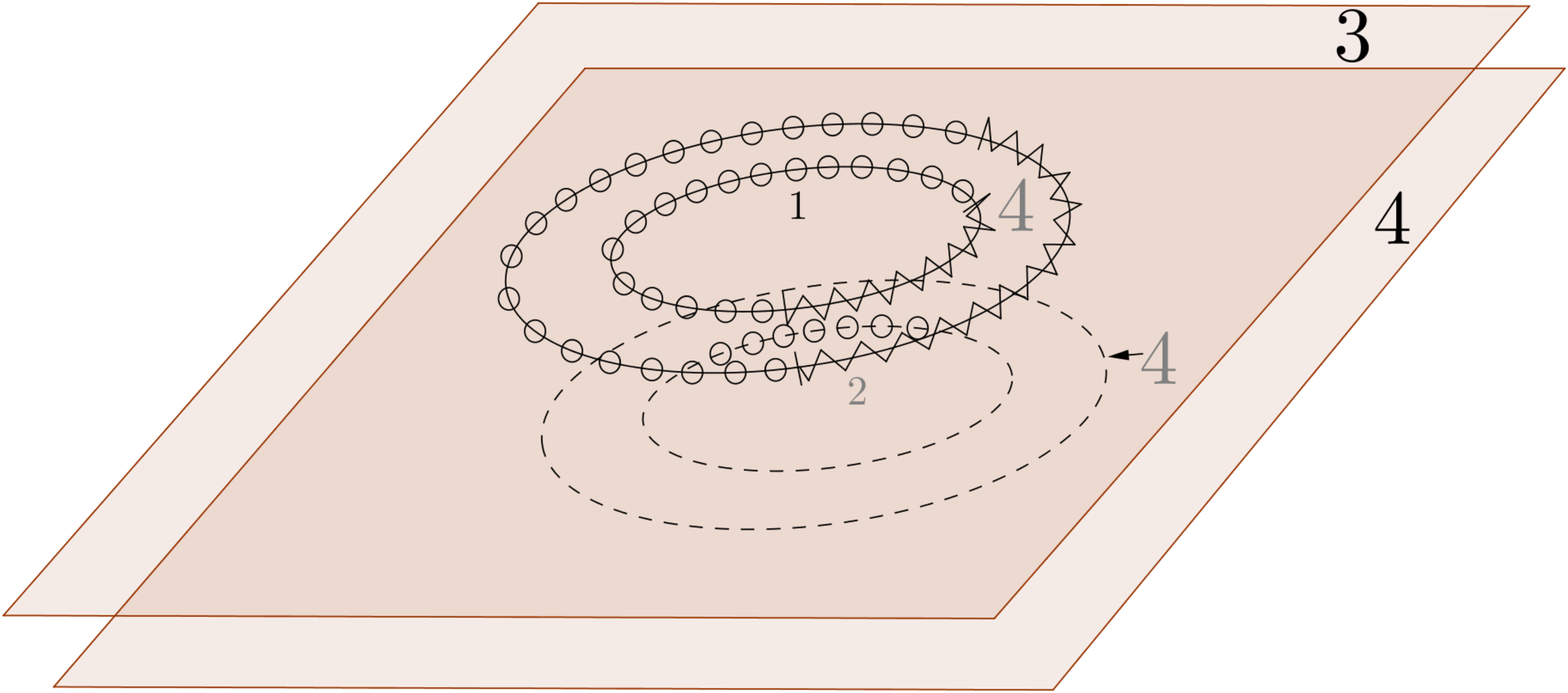}
\end{center}
\caption{}
Fig.Caption:Here we have drawn the markings ${\includegraphics[clip,width=1cm]{ring.eps}}$, and ${\includegraphics[clip,width=1cm]{wavy.line.eps}}$ along the edge of the complement3 and along the pieces of edge for respectely  ${\includegraphics[clip,width=1cm]{ring.eps}}$ on disc2 and ${\includegraphics[clip,width=1cm]{wavy.line.eps}}$ on disc1.
\end{figure}
\begin{figure}[H]
\begin{center}
\includegraphics[clip, width=15cm]{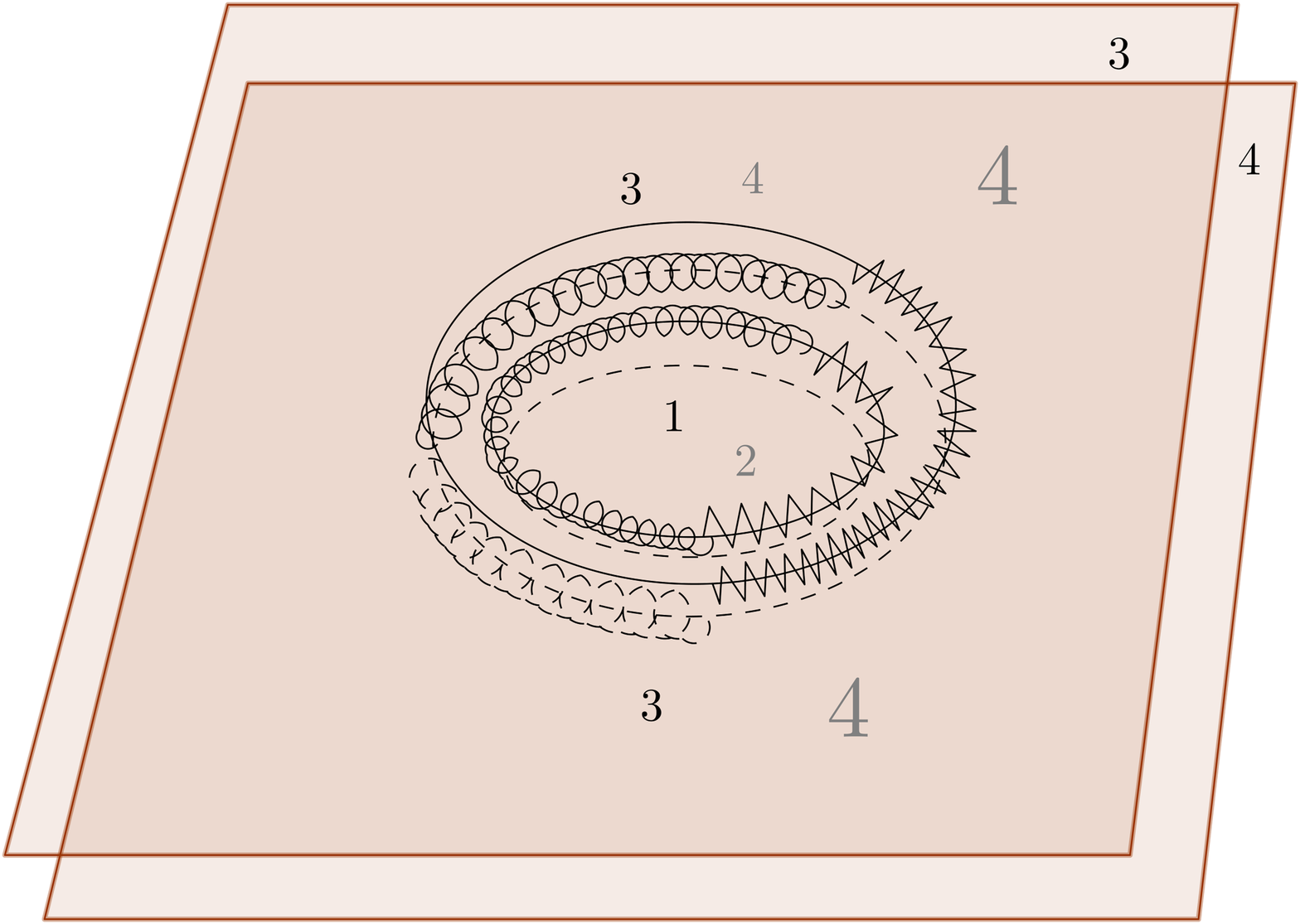}
\end{center}
\caption{}
Fig.Caption:On this fig.25 we show in a similar way the edge- decorations on the disc 1 marked ${\includegraphics[clip,width=1cm]{helix.eps}}$ and ${\includegraphics[clip,width=1cm]{wavy.line.eps}}$ and the corresponding pieces of the edges of the complements 3 and 4. 
\end{figure}

In this placement of these discs and complements of discs
one easily sees that by a simple cut running along the remaining
part of the unit circle, (a cut along which layer I is continued
into I\hspace{-1pt}I and oppositely,) we achieve to continue disc
for 1 into the complement of disc for 4 across the piece of
unit circle now marked  \includegraphics[clip, width=1cm]{helix.eps}.
Analogously the disc for 2 gets continued along the
 \includegraphics[clip, width=1cm]{ring.eps} edge across the cut
 into the complement of disc associated with string $3$.\\
 The cut connecting I with I\hspace{-1pt}I and oppositely must
 have a branch point in each of its two ends. About we manage to realize in the two-layered complex plane all
 the four gluings corresponding to objects in initial and
 final states being identified.\\
 
\begin{figure}[H]
\begin{center}
\includegraphics[clip, width=10cm]{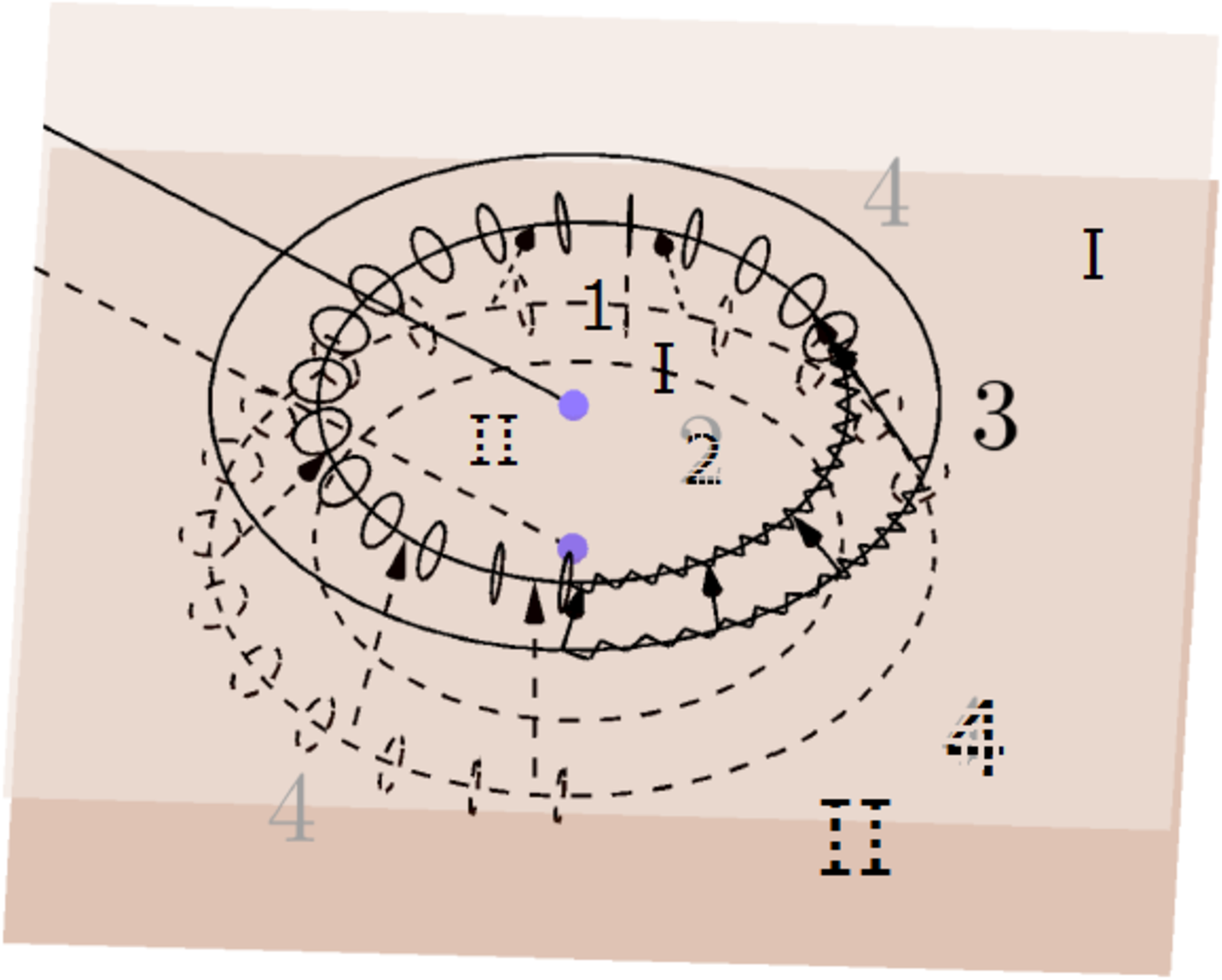}
\end{center}
\caption{}
Fig.Caption:On this figure we have drawn arrows to represent the gluings, but to avoid making the figure incomprehensively complicated we left out the arrows. That should have connected disc 2 with complement 3 by arrows crossing in between the arrows connecting disc 1 with complement 4 both marked ${\includegraphics[clip,width=1cm]{ring.eps}}$. Similarly we left out the arrows that should have connected disc 2 and complement 4 both in the second layer II. 
\end{figure} 
 
 The gluing together of the regions for the 
 functional along the curves where we identify objects  
in and outgoing states is supposed to lead to just the 
functional integral over the composed region (union of the
glued together regions.)\\
The full functional integrals with boundary
conditions from 26-momentum inlets in the
centers of the discs or in the infinity in the case 
of the complement of a disc becomes the overlap under
the specific way of identifying objects in the 
initial and the final state.\\
The full overlap is therefore a sum over all the ways of identifying the
objects of the overlaps under the specific identificaitons.\\
But we have made the approximation of only considering significant 
the identification patterns with the lowest number of jumps,
where neighboring objects do not follow
each other into the next cyclic chain.\\
Now we shall to evaluate the functional integral for a
given identification of objects make use of that 
modulo the anomaly -to which we shall return- the
value of the functional integral is invariant under conformal
transformations of the two-dimensional region associated with
this functional integral.\\
The \underline{most important step} in the calculation is now to by 
a conformal transformation map the double layered 
complex plane/or better Riemann sphere into a 
single layered one. Very suggestively this shall be
done by a square root type of analytic
function, because a square root ambiguity 
gives two possible values.\\
Then we can use a complex $z$ give two different
result values $f(z_{I})$ and $f(z_{II})$ depending
on which layer I or I\hspace{-1pt}I we imagine the complex number
$z$ to be.\\
Layer I drawn full \includegraphics[clip, width=1cm]{line.eps}\\
Layer I\hspace{-1pt}I drawn punctured  \includegraphics[clip, width=1cm]{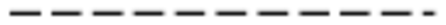}\\
All the four unit circles really coincide. It is only drawing technique they differ.
Edges of 1 and 3 to be identified \includegraphics[clip, width=1cm]{wavy.line.eps}
are on layer I.\\
Now we shall seek to map our two layered representation of the
two dimensional surface on which the functional integral is
to be calculated by an analytical/conformal map into a 
one-layered one. To get inspired to what map $f$ say to choose we
shall at first look at the situation near the two branch points.
Imagining a little ``circle'' meaning a curve in a distance $\epsilon$
($\epsilon$ small) from the very branch point we see that it get
actually the circumference $4\pi \epsilon$ rather as a
usual circle having only $2\pi \epsilon$.\\
This implies that the function performing the map
$f$ should have at square root singularity a square root like 
behavior near these two branch points.\\
Denoting the variable in the complex plane $z$ we get to each complex
number $z$ two sheets with the two points on the full
two-layered surface $z_{I}$ and $z_{II}$. 
Here the $z_{I}$ is the point on sheet I and 
$z_{II}$ the one on sheet II.\\
In this notation we can write down the attachment points
$a_{i}$ ($i=1,2,3,4$) for the four external points:
\begin{eqnarray}
a_{1}&=&O_{I}\\ \nonumber
a_{2}&=&O_{II}\\ \nonumber
a_{3}&=&\infty_{I}\\ \nonumber
a_{4}&=&\infty_{II}. 
\end{eqnarray}
Denoting the $z$-numbers corresponding to the branch points as 
$b_A$ and $b_B$ the need for square root behavior means
that near these branch points we shall have
\begin{equation}
f \stackrel{\sim}{\propto}\pm\sqrt{z-b_{i}} \hspace{1cm} i=A,B  
\end{equation}
or
\begin{equation}
f \buildrel{\sim}\over{\propto}\frac{1}{\pm \sqrt{z-b_{i}}}
\end{equation}
Here as well as when we finally choose $f$ it is to be
understood that the sign ambiguity is to be resolved
differently on the two sheets I and II
(for the same $z$). At other places singularities of
$f$ should rather be avoided. Actually the reader
may easily see that the function
\begin{equation}
f(z)=\pm\sqrt{\frac{z-b_{A}}{z-b_{B}}}
\end{equation}
is a good proposal for the transformation and
that with appropriate sheet dependence of the $\pm$
we get indeed a mapping into a single layer all
covering Riemann sphere(being the image of $f$).
Indeed we could check that $f$ maps the two layered
twodimensional region (for the functional integral)
surjectively (i.e. onto) to the Riemann sphere
(with variable $f$) by constructing the inverse
i.e. z as function of $f$:
This inverse is gotten by the equation
\begin{equation}
f^{2}=\frac{z-b_{A}}{z-b_{B}}
\end{equation}
meaning 
\begin{equation}
(b_{B}-z)f^{2}=z-b_{A} \hspace{0.5cm}
\Longrightarrow \hspace{0.5cm}
z=\frac{b_{B}f^{2}+b_{A}}{1+f^{2}}
\end{equation}
Under this map $f$ it is easily seen that the attachment
points $a_{1},a_{2},a_{3}$ and $a_{4}$ are mapped into
\begin{eqnarray}
f(a_{1})&=&\sqrt{\frac{b_{A}}{b_{B}}};\ f(a_{3})=1\\ \nonumber
f(a_{2})&=&\sqrt{\frac{b_{A}}{b_{B}}};\ f(a_{4})=-1
\end{eqnarray}
Remembering that the branch points lie on the unit circle in the $z$-plane
\begin{equation}
\mid b_{A}\mid^{2}=\mid b_{B}\mid^{2}=1
\end{equation}
we see that indeed all four images of the attachment points are also on the
unit circle (but now in the $f$-plane)
\begin{equation}
\mid f(a_{i})\mid^{2}=1\ {\rm for} i=1,2,3,4,
\end{equation}
Note immediately that such positions are just like Koba-Nielsen variables \cite{31}
for a four point Veneziano amplitude.\\

\subsection{The philosophy of counting}
The major purpose of these rather detailed calculations is to 
obtain the Veneziano amplitude correctly not only by having the
right external momentum dependence but also has the correct form
with respect to $26$-momentum independent factors, only depending on the 
integration dummy. In order for such an ambition level to make sense
we must start also from a well defined integration measure or rather we 
prefer to start from a well defined summation over a number specifying
how many objects are common for some couple of external strings.
But such a summation over a number of objects we have already developed above in section 5.1 especially(5.14).\\
We have a little freedom not yet used to orient the discs and the 
complements, which we can use to arrange that the two branch points
$b_{A}$, and $b_{B}$ become each others complex conjugate.
In fact we may take 
\begin{equation}
b_{A}=e^{i \delta};b_{B}=e^{-i \delta}
\end{equation}
with $\delta$ being a real angle. it is then easily seen that the
number say
\begin{equation}
N_{\includegraphics[clip, width=1cm]{wavy.line.eps}}\propto \delta
\end{equation}
so that the counting measure
$\displaystyle{\sum_{N_{\includegraphics[clip, width=1cm]{wavy.line.eps}}}}\sim dN_{\includegraphics[clip, width=1cm]{wavy.line.eps}}$
becomes proportional to the integral over this angle $\delta$
\begin{equation}
\displaystyle{\sum_{N_{\includegraphics[clip, width=1cm]{wavy.line.eps}}}}\sim dN_{\includegraphics[clip, width=1cm]{wavy.line.eps}}
\propto d\delta.
\end{equation}
\section{Evaluation of the integrand}
\subsection{Evaluation of the integrand}
Coming to the actual evaluation we must first face
the problem that obtaining the wave function
by a huge imaginary propagation leaves us
with an arbitrary and divergent normalization.\\
This divergence pops up by our need to led in at the
attachment of the external particles/strings via
small $\epsilon$-radius discs.\\
Then the divergence shows up by the functional integral
coming out in first approximation as the exponent of the 
classical action getting terms proportional to log $\epsilon$
into this classical action. However, since we shall not be
so ambitious as to calculate the absolute normalization of
the Veneziano amplitude it should be enough to just 
keep this $\epsilon$ cut off the same for all the
contribution to the Veneziano amplitude.
Especially we should keep our $\epsilon$'s for the four
external particle attachments fixed under the summation
over the different numbers of objects exchanged say
between 1 and 4. That is to say we must keep the 
$\epsilon$'s constant while varying the integration 
variable $\delta$. The terms proportional to the
log $\epsilon_{i}\ (i=1,2,3,4)$ are a priori
expected to be also proportional to 
$p^{\mu}_{i}p_{\mu i}$, whereas terms involving
the inner products of different external 26-momenta will
be convergent.\\
Taking these squares $p^{\mu}_{i}p_{\mu i}$ to be just
the masses squared
\begin{equation}
p^{\mu}_{i}p_{\mu i}=m^{2}_{i}
\end{equation}
and using that the mass of a mass-eigenvalue of a
string state is just a constant, we see that these divergences
$\propto log\ \epsilon_{i}$ are in principle not so severe
because they only give constant factors to the over all
amplitude, which we anyway give up calculating.\\
If we want to get the formula for the
$B\left(-\alpha(t),-\alpha(u)\right)$ written in terms of
the usual integration variable
\begin{eqnarray}
B\left(-\alpha(t),-\alpha(u)\right)=\int^{1}_{0}X^{-\alpha(u)-1}(1-X)^{-\alpha(t)-1}dX
\end{eqnarray}
If we identify the channel 
$1+\bar{4}\to\bar{2}+3$ as the $t$-channel and
$1+\bar{3}\to\bar{2}+4$ as the $u$-channel, while 
$1+2\to 3+4$ is the $s$-channel, then the integration 
variable for this expression
$\int X^{-\alpha(u)-1}(1-X)^{-\alpha(t)-1}dX$
shall be the anharmonic ratio that goes to zero in the situation
when the $u$-channel incoming $z$-variables approach each other.
In fact we must take
\begin{equation}
X=\frac{(z_{1}-z_{3})(z_{2}-z_{4})}{(z_{4}-z_{3})(z_{2}-z_{1})}
=\frac{z_{1}-z_{3}}{z_{4}-z_{3}}:\frac{z_{1}-z_{2}}{z_{4}-z_{2}}
\end{equation}
where we have chosen the denominator so as to $X\to 1$ when the
$z$-corresponding to say the incoming strings
$1+\bar{4}$ approach each other.\\
In the analogue model 
terminology we imagine
currents proportional to the 26-momentum
$p^{\mu}_{i}$ to be pumped in an $\epsilon$-disc at the
point $z_{i}$.\\
The external $26$-momentum conservation will allow these currents to 
flow in a conserved way. The current running in at
$z_{i}$ will if it just runs to infinity symmetrically by the in
the model assumed specific resistance
$2\pi\alpha^{\prime}$ (we shall use $2\pi\alpha^{\prime}$ rather
than $\pi\alpha^{\prime}$ because we work with the double lead
to a potential at the position $z$ in the Riemann sphere
$\frac{1}{2\pi}\cdot 2\pi\alpha^{\prime}ln\mid z-z_{i}\mid=\frac{\alpha^{\prime}}{2}ln\mid z-z_{i}\mid$.
Using this the total energy production rate in this
analogue model 
 would be
\begin{eqnarray}
\frac{1}{2}\sum_{\begin{array}{c}i,j\\i\neq j\end{array}}p^{\mu}_{i}p_{j\mu}ln\mid z_{j}-z_{i}\mid \\ \nonumber
({\rm the}\ j=i\ {\rm term\ diverge}) 
\end{eqnarray}
and so the exponential of this ``heat production rate'' becomes
\begin{equation}
\prod_{i,j}\left(\frac{\mid z_{j}-z_{i}\mid}{\epsilon_{i}}\right)^{-\alpha^{\prime}p^{\mu}_{i}p_{j\mu}}\cdot\prod N_{i}
\end{equation}
But these divergent factors although constant as functions of the external
momenta are not constant as function of the $z_{i}$-variables of say as function
of the anharmonic ratio $X$.\\
Remembering that the normalization factor for the
$i$th external particle is
\begin{equation}
N_{i}=\epsilon^{\alpha^{\prime}m^{2}_{i}}=\epsilon^{\alpha^{\prime}p^{2}_{i}}
\end{equation}
and the 26-momentum conservation
\begin{equation}
p^{\mu}_{1}+p^{\mu}_{2}=p^{\mu}_{3}+p^{\mu}_{4}
\end{equation}
or in an all ingoing notation
\begin{equation}
p^{\mu}_{1}+p^{\mu}_{2}+p^{\mu}_{3}+p^{\mu}_{4}=0
\end{equation}
we recognize that $\epsilon_{i}$ just
appears to the power 
\begin{equation}
``{\rm power\ of}\ \epsilon_{i}"=-\alpha^{\prime}p^{\mu}_{i}\sum_{j\neq i}p_{j\mu}+\alpha^{\prime}p^{2}_{i}=0
\end{equation}
and thus there is really no dependence on these
cut off $\epsilon_{i}$.\\
From the definition
\begin{equation}
X=\frac{z_{1}-z_{3}}{z_{4}-z_{3}}:\frac{z_{1}-z_{2}}{z_{4}-z_{2}}
\end{equation}
and we get
\begin{eqnarray}
1-X&=&\frac{(z_{4}-z_{3})(z_{2}-z_{1})-(z_{1}-z_{3})(z_{2}-z_{4})}{(z_{4}-z_{3})(z_{2}-z_{1})}\\ \nonumber
&=&\frac{z_{4}z_{2}+z_{3}z_{1}-z_{1}z_{2}-z_{3}z_{4}}{(z_{4}-z_{3})(z_{2}-z_{1})}\\ \nonumber
&=&\frac{(z_{3}-z_{2})(z_{1}-z_{4})}{(z_{4}-z_{3})(z_{2}-z_{1})}.
\end{eqnarray}
Using in the $p_{1}+p_{2}+p_{3}+p_{4}=0$ notation
\begin{eqnarray}
s&=&m^{2}_{1}+m^{2}_{2}+2p_{1}\cdot p_{2}\\ \nonumber
&=&m^{2}_{3}+m^{2}_{4}+2p_{3}\cdot p_{4}\\ \nonumber
t&=&m^{2}_{1}+m^{2}_{4}+2p_{1}\cdot p_{4}\\ \nonumber
&=&m^{2}_{2}+m^{2}_{3}+2p_{3}\cdot p_{2}\\ \nonumber
u&=&m^{2}_{1}+m^{2}_{3}+2p_{1}\cdot p_{3}\\ \nonumber
&=&m^{2}_{2}+m^{2}_{4}+2p_{2}\cdot p_{4}
\end{eqnarray}
and thus
\begin{equation}
s+t+u=m^{2}_{1}+m^{2}_{2}+m^{2}_{3}+m^{2}_{4}
\end{equation}
to replace
\begin{equation}
p_{1}\cdot p_{2}=\frac{1}{2}(m^{2}_{3}+m^{2}_{4}-t-u).
\end{equation}
We obtain
\begin{eqnarray}
&&\left(\Pi N_{i}\right)\cdot\prod_{i,j}\Biggl(\frac{\mid z_{i}-z_{j}\mid}{\epsilon_{i}}\Biggr)^{-\alpha^{\prime}p^{u}_{i}\cdot p_{ju}}\\ \nonumber
&=&\prod_{\begin{array}{c}i,j\\i\neq j\end{array}}\mid z_{i}-z_{j}\mid^{-\alpha^{\prime}p^{u}_{i}\cdot p_{ju}}
\cdot\biggl(\prod_{\begin{array}{c}i,j\\i\neq j\end{array}}\mid z_{i}-z_{j}\mid^{-\alpha^{\prime}m^{2}_{i}}\biggr)\\ \nonumber
&=&\prod_{\begin{array}{c}(i,j)\\{\rm with}\ i\neq j\ {\rm but\ each}\ p\end{array}}
\mid z_{i}-z_{j}\mid^{-2\alpha^{\prime}p^{u}_{i}p_{ju}-\alpha^{\prime}(m^{2}_{i}+m^{2}_{j})}\\ \nonumber
&=&\prod_{\begin{array}{c}(i,j)\\{\rm on\ one\ order}\end{array}}\mid z_{i}-z_{j}\mid^{\alpha^{\prime}(p_{i}-p_{j})^{2}}
\end{eqnarray}
\begin{align}
\cdot&\mid z_{1}-z_{3}\mid^{-\alpha^{\prime}u}\\ \nonumber
\cdot&\mid z_{2}-z_{4}\mid^{-\alpha^{\prime}u}\\ \nonumber
\cdot&\mid z_{1}-z_{2}\mid^{-\alpha^{\prime}(-t-u+m^{2}_{3}+m^{2}_{4})}\\ \nonumber
\cdot&\mid z_{3}-z_{4}\mid^{-\alpha^{\prime}(-t-u+m^{2}_{1}+m^{2}_{2})}\\ \nonumber
\cdot&\mid z_{1}-z_{4}\mid^{-\alpha^{\prime}t}\\ \nonumber
\cdot&\mid z_{2}-z_{3}\mid^{-\alpha^{\prime}t}\\ \nonumber
=&X^{-\alpha^{\prime}u}\left(1-X\right)^{-\alpha^{\prime}t}
\mid z_{3}-z_{4}\mid^{-\alpha^{\prime}(m^{2}_{1}+m^{2}_{2})}
\cdot\mid z_{1}-z_{2}\mid^{-\alpha^{\prime}(m^{2}_{3}+m^{2}_{4})}
\end{align}
%
\subsection{What is required  to finish Veneziano model?}
To get the last bit of the way to obtain all three terms in the four point Veneziano
model in our object scheme we have to obtain a definite counting of the number
of ways of identifying or better bringing in correspondence the objects in the
initial state with those in the finial state.\\
In our previous article we used as a combined gauge or parameter $\tau_{R}$ choice 
and discretization to impose the condition that each object has it $J^{+}$,
a special component of its $J^{\mu}$ take a specific value
\begin{equation}
J^{+}=\frac{a\alpha^{\prime}}{2}
\end{equation}
for all the objects.\\
This ``longitudinal momentum'' (or ``longitudinal $J^{\mu}$'') $J^{+}$
(which is essentially the $+$ component of momentum
of the object) is defined
\begin{equation}
J^{+}=J^{0}+J^{25}
\end{equation}
where $J^{25}$ is the infinite momentum frame direction. If, and that
is indeed true since classically
\begin{equation}
(J^{\mu})^{2}=0,
\end{equation}
the $J^{\mu}$ is ``on shell'' as a momentum then
\begin{equation}
J^{+}\ge 0.
\end{equation}
But if we allow ``energy'' $J^{0}$ to be negative then also
$J^{+}$ is not guaranteed to be positive.\\
As long as the $J^{+}$'s are guaranteed positive as we
used in previous paper, then one can devide the positive
$+$-momentum into $J^{+}$'s all being some given positive number
\begin{equation}
J^{+}(I)=\frac{a\alpha^{\prime}}{2}.
\end{equation}
But now in order to be allowed to hope for obtaining all the
3 terms in the full Veneziano amplitude we must accept negative
$J^{+}$'s.\\
So we must choose a gauge or parametrization and discretization
choice that is more liberal with respect to allowing negative $J^{+}$ too.
The obvious suggestion is that in constructing the dicretization
we first imagine dividing the cyclic chians corresponding to the strings
into pieces with negative $J^{+}$'s and  pieces with positive $J^{+}$'s.
In order that this shall be nice it should be so that the $\dot{X}^{\mu}_{R}(\tau_{R})$
that is proportional/essentially the same as $J^{\mu}$ is so smooth
that the sign of $\dot{X}^{+}_{R}(\tau_{R})$
\begin{equation}
sign\{\dot{X}^{+}_{R}(\tau_{R})\}
\end{equation}
is constant over intervals of reasonably large size.\\
With some -may be a bit vague- continuity assumption these
large intervals of fixed signs $sign\{\dot{X}^{+}_{R}(\tau_{R})\}$
is justified. Also one has in this spirit also the assumption 
that the pieces of a cyclic chain in an initial state string
going in the overlap into a given final state cyclic chain will
consist dominantly of very few pieces. It shall with highest weight
be connected to one connected piece only. In the scattering the
negative energy $J^{0}$ or negative $J^{+}$ pieces have to for say
an initial cyclic chain must either annihilate with positive piece in another 
incoming cyclic chain or go on as a negative piece in the final state.\\

\section{Dominant term in $1+2\to 3+4$ scattering}
\subsection{Dominant term in $1+2\to 3+4$ scattering}
\ \ \ \ Mandelstam has diagrams describing his contribution to the Veneziano amplitude in which two strings are interacting by their ends (contrary to ours which rather interact the intuitively most likely way by crossing each other on a random point somewhere on the string ) and then t propagates them in what is effectively imaginary time. Some such imaginary time before they again split into two strings. It we now say that hidden in this imaginary time propagation can be hitten a partial annihilation, it means that Mandelstam has something that very likely is partial annihilation of pieces of one of the incoming strings with part of the other one. Translated to our cyclically ordered chain that could easily mean that Mandelstam has indeed in his picture what we take as the effect of having negative $P^+$.
To say it shorter: Mandelstam has, in real physical sense, involved imaginary time development – while we only use it as a trick to make wave functions. Thus he also has this physically very strange thing that the strings interact with their END-points rather than what classically thinking is definitely more likely by the random points of the two incoming strings meeting. 
This means that he -Mandelstam- has a strange quantum effect treated by a very formal treatment with imaginary time included. So from our “more realistic” point of view Mandelstam, and most string theorists work with some quantum tunneling as if it were a true physical description, and then sometimes one might get possibility for doing something – like getting all three terms-so that it looks physically done, but in reality might not be possible (with the assumed positive $P^+$ ).
Further negative $P^+$ case will be investigated in our forthcoming paper. 

According to our somewhat vague continuity assumption the dominant term
to a scattering amplitude should come from a system of houw pieces of the incoming cyclic
chains going into the final state or annihilate in the most simple way.
I.e. it is the system with fewest pieces involved. Let us see what is this dominant term in
the case of one of the initial cyclic chains having a ``negative'' piece:

\begin{figure}[htbp]
\begin{center}
\includegraphics[clip,width=5cm]{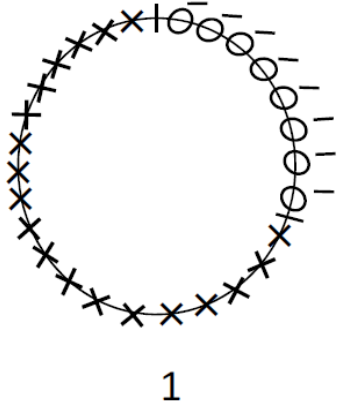} 
\end{center}
\caption{Dominant term in the case of the cyclic chains
with a negative piece}
\end{figure}

We put minuses to indicate the ``negativeness'' of a piece of the
to string $1$ correspinding ``negative part'' in the string 1 here denoted\\

\begin{figure}[H]
\begin{center}
\includegraphics[clip, width=4cm]{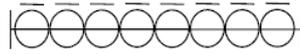} 
\end{center}
\caption{Negative part of pieces of string $1$}
\end{figure}

The interesting case is when the piece\
\includegraphics[clip,width=2cm]{negative.part.eps}\
of negative cyclic
chain in string $1$ gets annihilated by a correspoinding piece in
the string $2$ cyclic chaing.\\
Under the assumption of as few pieces as possible the
cyclic chain for string $2$ should only be split into the two parts: 
one annihilating with the\ \mbox{\includegraphics[clip,width=2cm]{negative.part.eps}}\
from cyclic chain $1$ and one part continuing to just
one of the final state strings, say string $3$, or rather its cyclic chain.
So let us write the whole initial state cyclic chain as:

\begin{figure}[htbp]
\begin{center}
\includegraphics[clip,width=3cm]{cyclic.chain.1.eps}
\hspace{1.5cm}
\includegraphics[clip,width=3cm]{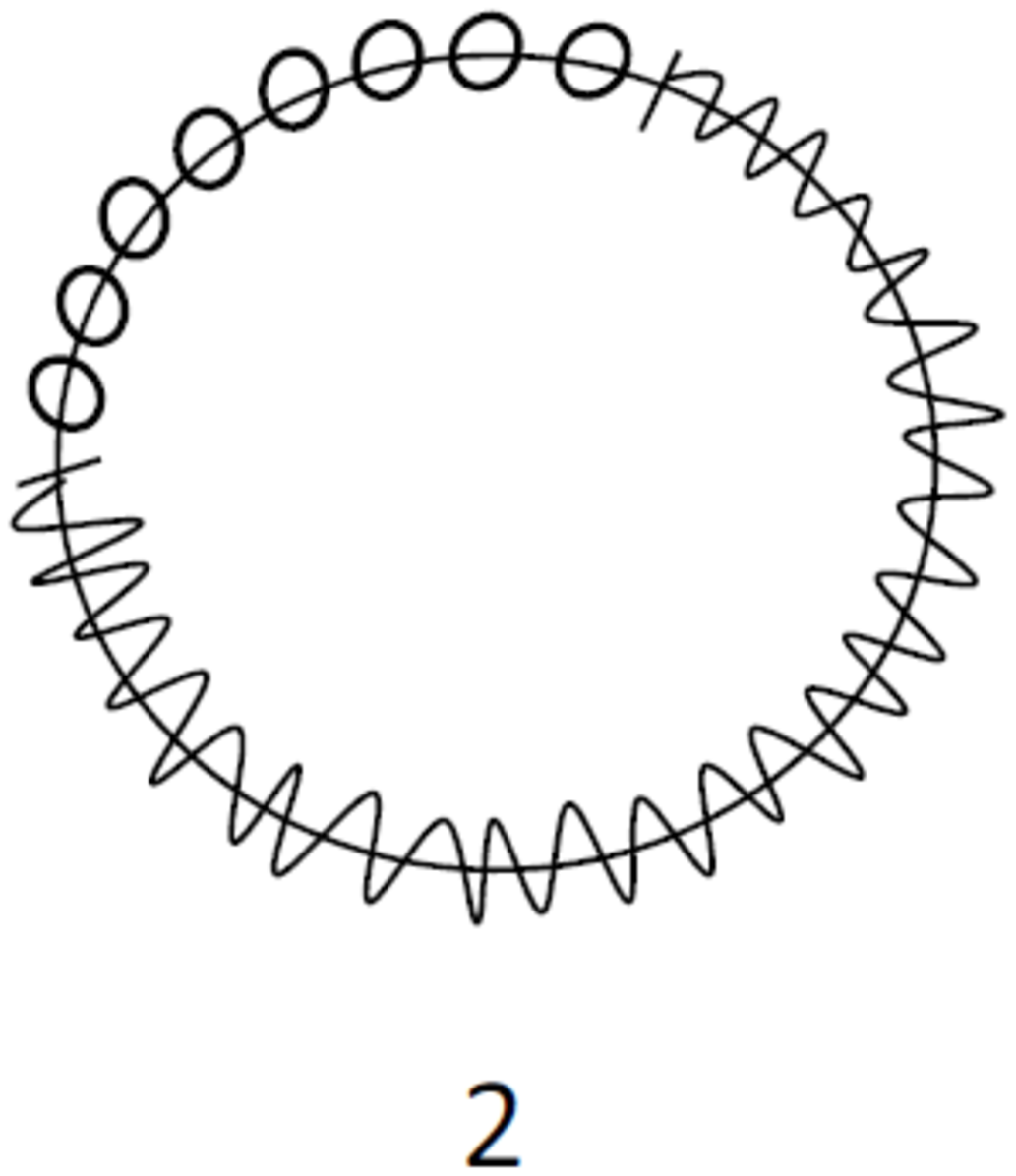} 
\end{center}
\caption{Incoming string $1$ cyclic chain negative part
and string $2$ cyclic chain splits into the two parts:
one annihilates with the negative piece from the 
cyclic chain $1$ and one part of the final string
e.g. string $3$ cyclic chain}
\end{figure}
 
Now we have in the final state for the considered process
\begin{equation}
1+2\to 3+4
\end{equation}
also two open strings and therefore two cyclic chains $3$ and $4$.\\
From the continuity assumption or assumption of fewest pieces dominating 
then all the\ \includegraphics[clip,width=2cm]{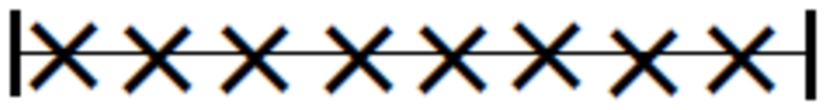}\ goes to one, say cyclic chain $4$, while the piece from $2$\
\includegraphics[clip,width=2cm]{wavy.line.eps}\ continues in cyclic chain $3$.\\
Now the only way to have the rests of $3$ and $4$ dispences which is to let a piece
from $3$ and from $4$ annihilate each other.\\
Finally we thus end up with the scheme represented by the figure 30.

\begin{figure}[htbp] 
\begin{center}
\includegraphics[clip,width=10cm]{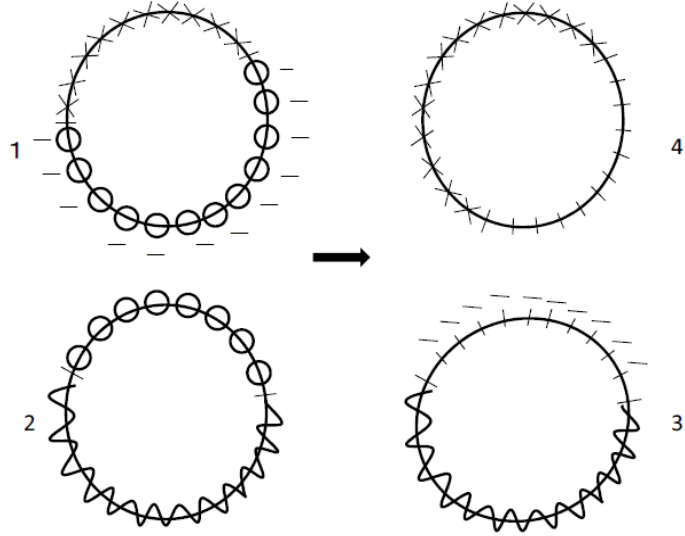}
\end{center}
\caption{The analogue of fig. 19 but now negative $P^+$ (and thus energy) on the with – ‘s marked pieces of the cyclic chains for string 1 and string 3.}
\end{figure}

The ``gauge or discretization condition'' that we here choose is proposed:\\
\begin{equation}
{\rm For\ positive\ pieces:}\hspace{0.5cm}
J^{+}=\frac{a\alpha^{\prime}}{2}
\end{equation}
\begin{equation}
{\rm For\ negative\ pieces:}\hspace{0.5cm}
J^{+}=-\frac{a\alpha^{\prime}}{2}.
\end{equation}
Then conservation of the $p^{+}$-component of the $26$-momentum implies
that the number of (even) objects in the positive pieces of the
initial state minus those in the negative pieces, in our example
\begin{equation}
N^{(i)}_{\includegraphics[clip,width=1cm]{xxx.eps}}+N^{(i)}_{\includegraphics[clip,width=1cm]{wavy.line.eps}}+
N^{(i)}_{\includegraphics[clip,width=1cm]{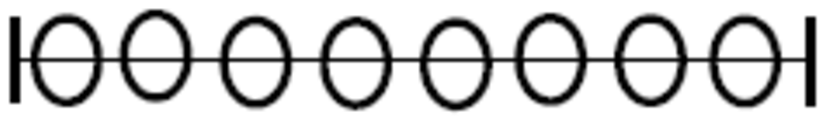}}-
N^{(i)}_{\includegraphics[clip,width=1cm]{negative.part.eps}}
\end{equation}
must be equal to the same quatity for the final state
\begin{equation}
N^{(f)}_{\includegraphics[clip,width=1cm]{xxx.eps}}+N^{(f)}_{\includegraphics[clip,width=1cm]{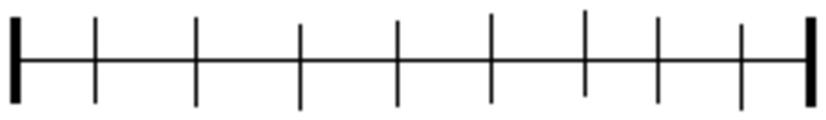}}+
N^{(f)}_{\includegraphics[clip,width=1cm]{wavy.line.eps}}-
N^{(f)}_{\includegraphics[clip,width=1cm]{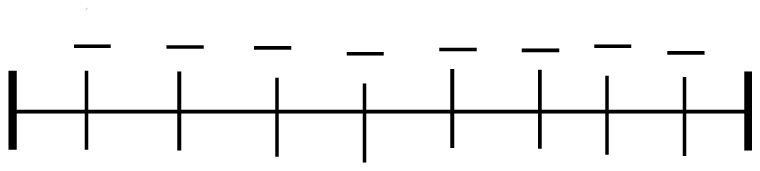}}
\end{equation}
That is to say we must have
\begin{eqnarray}
\label{eqNiNf}
N^{(f)}_{\includegraphics[clip,width=1cm]{xxx.eps}}+N^{(f)}_{\includegraphics[clip,width=1cm]{4.eps}}+
N^{(f)}_{\includegraphics[clip,width=1cm]{wavy.line.eps}}-
N^{(f)}_{\includegraphics[clip,width=1cm]{3.eps}}\\ \nonumber
=
N^{(i)}_{\includegraphics[clip,width=1cm]{xxx.eps}}+N^{(i)}_{\includegraphics[clip,width=1cm]{wavy.line.eps}}+
N^{(i)}_{\includegraphics[clip,width=1cm]{ooo.eps}}-
N^{(i)}_{\includegraphics[clip,width=1cm]{negative.part.eps}}. 
\end{eqnarray}
%
But noticing that in order to have an ``annihilation'' of a series of 
positive energy objects with a series of negative energy ones
there should be equally many of them, we have
\begin{equation}
N^{(f)}_{\includegraphics[clip,width=1cm]{4.eps}}=N^{(f)}_{\includegraphics[clip,width=1cm]{3.eps}}
\end{equation}
and
\begin{equation}
N^{(i)}_{\includegraphics[clip,width=1cm]{ooo.eps}}=N^{(i)}_{\includegraphics[clip,width=1cm]{negative.part.eps}}
\end{equation}
and that also for the correspondence of the objects in initial state
with those in the final state requires
\begin{equation}
N^{(f)}_{\includegraphics[clip,width=1cm]{xxx.eps}}=N^{(i)}_{\includegraphics[clip,width=1cm]{xxx.eps}}
\end{equation}
and
\begin{equation} 
N^{(f)}_{\includegraphics[clip,width=1cm]{wavy.line.eps}}=N^{(i)}_{\includegraphics[clip,width=1cm]{wavy.line.eps}}
\end{equation}
We see that the equation (\ref{eqNiNf}) is trivially satisfied.\\
Thinking of the important part of the set of the cyclically ordered chains
of objects which really exists,\\

\begin{figure}[H]
\begin{center}
\includegraphics[clip, width=6cm]{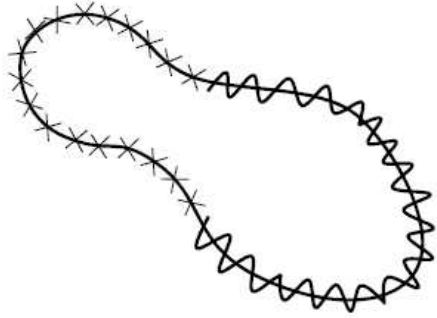} 
\end{center}
\caption{The important part of the set of the cyclically
ordered chain of objects that really exists on the
figure depicted as Figure \ref{svms.3}}
\label{svms.3}
\end{figure}

The quantity that has to be summed over
is the amount of objects really existing which belong to
string $1$ or equivalently string $4$ \includegraphics[clip,width=2cm]{xxx.eps} 
as contrary to the rest which belongs to string $2$ or equivalently to
string $3$ \includegraphics[clip, width=2cm]{wavy.line.eps}. 
The sum of the numbers of objects in these two classes, 
the $1$ or $4$ \includegraphics[clip,width=2cm]{xxx.eps} and the
$2$ or $3$ \includegraphics[clip, width=2cm]{wavy.line.eps} is
constant for given external $26$ dimensional-momenta of the four strings
\begin{equation}
N^{(i\ {\rm or}\ f)}_{\includegraphics[clip,width=1cm]{wavy.line.eps}}+N^{(i\ {\rm or}\ f)}_{\includegraphics[clip,width=1cm]{4.eps}}
\stackrel{\propto}{=} P^{+}_{total}
\end{equation}
due to that this sum is proportional to the
sum of the ``longitudinal'' momenta $p^{+}_{1}$ and $p^{+}_{2}$
of string $1$ and string $2$.  So for a set of fixed external momenta
for which we want to evaluate the overlap or the scattering amplitude
the only physically existing variation in the numbers of
objects in the different
%
ways of splitting the already fixed sum
$N_{\includegraphics[clip,width=1cm]{wavy.line.eps}}+N_{\includegraphics[clip,width=1cm]{4.eps}}$
into its two parts. Each choice of a part should be weighted with no extra weight, when we just
compare one positive integer value for $N_{\includegraphics[clip,width=1cm]{wavy.line.eps}}$
with another positive integer value for $N_{\includegraphics[clip,width=1cm]{wavy.line.eps}}$.\\

\subsection{Problems in getting the weight of Veneziano integral easily}
To make the derivation of the Veneziano model from our 
object-based string field theory so easy as possible while 
still using the ``gauge'' in which
$J^{+}=\pm \frac{a \alpha^{\prime}}{2}$, which could be called
IMF($=$ infinite momentum frame) gauge, we shall for each of the three
terms which we hope to obtain in full Veneziano model
choose a different condition specifying the Lorentz frame shall use.\\

%
We want to arrange that when say
there are no more objects left on string $1$ we truly reached the end 
of the chain of possibilities to be summed over. 
Essentially we want the reach of zero or run out of objects to
occur ``simultaneously'' for say both string $1$ and 
some other string that runs out.\\
Let us be more precise and first notice as one varies the numbers
of objects $N_{\includegraphics[clip,width=1cm]{wavy.line.eps}}$,
$N_{\includegraphics[clip,width=1cm]{ooo.eps}}$,
$N_{\includegraphics[clip,width=1cm]{xxx.eps}}$, and
$N_{\includegraphics[clip,width=1cm]{4.eps}}$ while
keeping the external $P^{+}$-momenta fixed, a set of small variations
$\Delta N_{\includegraphics[clip,width=1cm]{wavy.line.eps}}$,
$\Delta N_{\includegraphics[clip,width=1cm]{ooo.eps}}$,
$\Delta N_{\includegraphics[clip,width=1cm]{xxx.eps}}$, and
$\Delta N_{\includegraphics[clip,width=1cm]{4.eps}}$ of 
these quantities must obey
\begin{equation}
\Delta N_{\includegraphics[clip,width=1cm]{wavy.line.eps}}=
-\Delta N_{\includegraphics[clip,width=1cm]{ooo.eps}}=
-\Delta N_{\includegraphics[clip,width=1cm]{xxx.eps}}=
\Delta N_{\includegraphics[clip,width=1cm]{4.eps}}.
\end{equation}
For fixed external $P^{+}$'s we has that
\begin{eqnarray}
N_{\includegraphics[clip,width=1cm]{xxx.eps}}-N_{\includegraphics[clip,width=1cm]{ooo.eps}}=const_{1}\ {\rm (from\ str.\ 1)} \\ \nonumber
N_{\includegraphics[clip,width=1cm]{ooo.eps}}+N_{\includegraphics[clip,width=1cm]{wavy.line.eps}}=const_{2}\ {\rm (from\ str.\ 2)} \\ \nonumber
N_{\includegraphics[clip,width=1cm]{wavy.line.eps}}-N_{\includegraphics[clip,width=1cm]{4.eps}}=const_{3}\ {\rm (from\ str.\ 3)} \\ \nonumber
N_{\includegraphics[clip,width=1cm]{4.eps}}+N_{\includegraphics[clip,width=1cm]{xxx.eps}}=const_{4}\ {\rm (from\ str.\ 4)}
\end{eqnarray}
These numbers of objects, negative $J^{+}$ of positive $J^{+}$, must be positive and
thus they must run in the intervals
\begin{eqnarray}
0&\le& N_{\includegraphics[clip,width=1cm]{wavy.line.eps}}\le 4P^{+}_{2}/a \\ \nonumber
0&\le& N_{\includegraphics[clip,width=1cm]{4.eps}}\le 4P_{4}/a
\end{eqnarray}
Now we want to make the boundaries so simple as possible by
letting the ranges of the different $N$'s be the same, otherwise 
we get the problem that since only one
$N$ can be varied independently we could \underline{not}
fill out both ranges. Indeed we thus want to make the
two ranges correspond to each other and thus especially
need that the length of the interval for $N_{\includegraphics[clip,width=1cm]{wavy.line.eps}}$ 
which is $4P^{+}_{2}/a$ be the same as that for 
$N_{\includegraphics[clip,width=1cm]{ooo.eps}}$, namely $4P^{+}_{4}/a$.
Thus we are driven to -for simplicity- to claim that we shall
suggest to arrange by a Lorentz transformation that
\begin{equation}
P^{+}_{2}=P^{+}_{4}.
\end{equation}
If we decide that also $N_{\includegraphics[clip,width=1cm]{ooo.eps}}$ and
$N_{\includegraphics[clip,width=1cm]{4.eps}}$ shall run over the
same interval, we must let
$N_{\includegraphics[clip,width=1cm]{wavy.line.eps}}$ and
$N_{\includegraphics[clip,width=1cm]{4.eps}}$ start simultaneously at
zero and similarly also $N_{\includegraphics[clip,width=1cm]{ooo.eps}}$
and $N_{\includegraphics[clip,width=1cm]{xxx.eps}}$ should start at
zero simultaneously. So at the end we are driven towards
\begin{eqnarray}
N_{\includegraphics[clip,width=1cm]{ooo.eps}}&=&N_{\includegraphics[clip,width=1cm]{xxx.eps}}\in (0,4P_{2}/a) \\ \nonumber
N_{\includegraphics[clip,width=1cm]{wavy.line.eps}}&=&N_{\includegraphics[clip,width=1cm]{4.eps}}\in (0,4P_{2}/a),
\end{eqnarray}
these two couples running though opposite to each other.  The first of these two
equations imply that$P^{+}_{1}=0$ and the second one
that $P^{+}_{3}=0$. In conclusion we suggest that to avoid troubles with
not being allowed to sum over just
one interval we shall restrict our consideration to the
special situation achievable in principle a Lorentz
transformation such that 
\begin{eqnarray}
P^{+}_{1}&=&P^{+}_{3}=0\\ \nonumber
{\rm and}\hspace{1cm} P^{+}_{2}&=&P^{+}_{4}.
\end{eqnarray}
With this special type of external momentum
configuration we remark that we have in the
cyclic chains for string $1$ and string $3$ just
the \underline{same} number of negative and
positive $J^{+}$ objects.\\
On the other hand we take the cyclic chains of
strings $2$ and $4$ to have only positive$J^{+}$
objects.\\
Analogously to the writing of the wave function
in the case of only positive $J^{+}$ objects as
functional integrals, we shall also write
the wave functions here. For strings $2$ and $4$ where 
we have only positive $J^{+}$ objects, it is 
exactly as in the totally positive $J^{+}$ case.
But for strings $1$ and $3$, we have to ensure ourselves
that we can just say that looking for transverse 
momentum the functional integral is just
given by the number of objects no matter if they
have positive $J^{+}$ or negative $J^{+}$.
This should be so because in the functional integral formally the 
different  components of 26 momentum, i.e. different $\mu$, are
completely decoupled. \\
Now, however, we discovered a little problem -the species doubler problem \cite{30}
the herely associated nonorientation invariant
``continuity condition''.\\
If we shall indeed be able to have a piece of cyclic chain with
positive $J^{+}=\frac{a\alpha^{\prime}}{2}$ cancel a piece with
negative $J^{+}=-\frac{a\alpha^{\prime}}{2}$ then the
continuity conditions for the ``negative'' and for the
``positive'' pieces must match so as to make cancellation
possible. The requirement needed is that we can have a ``positive''
piece say with both odd and even $J^{\mu}$'s for some 
component index $\mu$ be just opposite to those for a
``negative'' piece. Now, however, because of the
non-orientation invariance it is very important how
we decide to order the numbering $I$ along the two pieces
that shoule cancel. To see how we
need to require the orientations let us consider
the following figure illustrating a cyclic chain in
the Minkowsky space time which is interpreted
as composition from two different cyclic chains,
one of which has a ``negative'' part:\\

\begin{figure}[htbp]
\begin{center}
\includegraphics[clip, width=6cm]{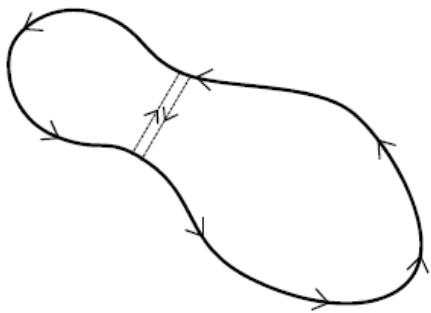} 
\end{center}
\caption{The figure shows how we need to require the 
orientations. The cyclic chain in the Minkowski space time.
There should be considered:
(a)There are negative and compensating parts indicated by dotted line. 
These parts are pair of compensating pieces.
(b)The cyclic chain with positive parts denoted
by solid line with positive parts.}
\end{figure}

Here the phantasy part for a pair of compensating pieces is denoted\\

where the orientation arrows have been put on so that
the total can also be interpreted as the two
separate cyclic chains\\
%
The point to be drawn from these figures is that provided we insist on
a simple cyclic ordering for our cyclic chains we must accept an 
\underline{opposite} 
ordering of the two conpensating pieces.
This in turn means that requiring for the ``positive'' pieces
the usual ``continuity condition'' we are driven to take the
opposite ``continuity condition'' for the ``negative'' pieces.
In other words we are driven to:\\
For the number I odd case:
\begin{eqnarray}
J^{\mu}(I+1)\approx -\Pi\alpha^{\prime}\left(\Pi^{\mu}(I+1)-\Pi^{\mu}(I-1)\right)\approx J^{\mu}(I-1)\\ \nonumber
{\rm for}\hspace{1cm} J^{+}(I)=\frac{a\alpha^{\prime}}{2} \hspace{1cm} {\rm i.e.\ positive}
\end{eqnarray}
and
\begin{eqnarray}
J^{\mu}(I+1)\approx +\Pi\alpha^{\prime}\left(\Pi^{\mu}(I+1)-\Pi^{\mu}(I-1)\right)\approx J^{\mu}(I-1)\\ \nonumber
{\rm for}\hspace{1cm} J^{+}(I)=-\frac{a\alpha^{\prime}}{2} \hspace{1cm} {\rm i.e.\ ``negative''}.
\end{eqnarray}
When going to the discussion using the complex plane the non-orientation 
invariant ``continuity condition'' can, as is rather easy, 
be considered a Caucy-Riemann condition for our $\phi^{\mu}$
in the functional integral. When we, as the technical trick, to 
produce the ground state via an imaginary $\tau$-time propagation
``sneak in'' conplex numbers our a priori real  $\phi^{\mu}$
gets complex as say classical solution.

In any case if we under conformal transformations want to 
keep the continuity conditions undisturbed, we must take
it that to the negative objects,
$J^{+}<0$, corresponding regions in the two dimentional
space for the functional integral have anti-analytic
rather than analytic $\phi^{\mu}$'s.
Here we imagine that we associate with each section 
of negative $J^{+}$ a behind it region in the 
two-dimensional region for the functional integral.
In this way we divide up the two-dimensional region
into pieces that should be conformally transformed
only under respectively analytical and antianalytical 
maps in order that the ``continuity conditons'' be kept.\\
At the boarder line between the ``negative'' and ``positive''
regions the regions goes from analytical to anti-analytical,
and a natural way to put it into the complex plane
would be to give it a reflexion line here between
the two regions.
In this way like the strings $2$ and $4$ with only ``positive'' objects 
get their wave functions represented by function on a region
being a half infinite cylinder or conformally equivalent disc
(with the objects sitting on the edge and the momentum of
the string entering in the centrum.)\\
Also the strings $1$ and $3$ get their wave functions
reresented by functional integral of what is represented as
a disc or half infinite cylinder.\\

\begin{figure}[H]
\begin{center}
\includegraphics[clip, width=5cm]{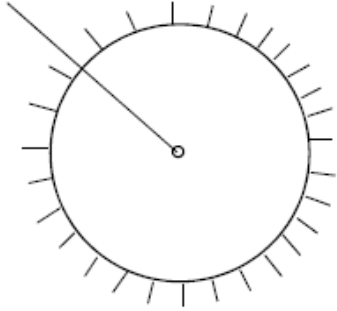} 
\end{center}
\caption{The wave functions for the strings $1$ and $3$
by the functional integral of following figure
as a disc or a half infinite cylinder.}
\label{svms.34}
\end{figure}

Now these 4 figures for the four external particles/strings
have to be glued together according to the rules of the
identification of the initial and final states in terms of
even object.\\

However, we can only glue together, to
either annihilate or identity -from initial and final state-
pieces of the edges of these discs provided the identified
or annihilated pieces have the same numbers of objects.\\
Now under the summation variation of the numbers 
$N_{\includegraphics[clip, width=1cm]{wavy.line.eps}}$,
$N_{\includegraphics[clip, width=1cm]{ooo.eps}}$,
$N_{\includegraphics[clip, width=1cm]{xxx.eps}}$, and
$N_{\includegraphics[clip, width=1cm]{4.eps}}$,
however the number of ``negative'' and 
``positive'' objects on the edge of the disc for
strings $1$ and $3$ varies.\\
So to bring e.g. the number  $N_{\includegraphics[clip, width=1cm]{ooo.eps}}$
of ``negative'' objects on the cyclic chain for string $1$ to annihilate
with the corresponding $N_{\includegraphics[clip, width=1cm]{ooo.eps}}$
``positive'' objects on string $2$ marked \includegraphics[clip, width=1cm]{ooo.eps}
in a way in which pieces of a unit circle match we must first make a 
conformal transformation of the disc for string $1$ so that its edge for
the objects marked \includegraphics[clip, width=1cm]{ooo.eps} 
instead of being -as on the a priori figure- $180^\circ$ rather becomes
$2\pi\frac{\includegraphics[clip, width=1cm]{ooo.eps}}{4P^{+}_{2}/a}$, namely
the angle corresponding to the \includegraphics[clip, width=1cm]{ooo.eps}
marking of the disc for string$2$.\\
This conformal map must be an exponentiation with the power
\begin{equation}
\frac{2N_{\includegraphics[clip, width=1cm]{ooo.eps}}}{4P^{+}_{2}/a}
=\frac{2N_{\includegraphics[clip, width=1cm]{ooo.eps}}}{N_{\includegraphics[clip, width=1cm]{ooo.eps}}+N_{\includegraphics[clip, width=1cm]{wavy.line.eps}}}.
\end{equation}
I.e. the variable say $y$ for the disc should be transformed
\begin{equation}
y\to \xi=y^{\frac{2N_{\includegraphics[clip, width=1cm]{ooo.eps}}}{N_{\includegraphics[clip, width=1cm]{ooo.eps}}+N_{\includegraphics[clip, width=1cm]{wavy.line.eps}}}}
\end{equation}
so that the unit circle parametrization by $\delta$ say as
\begin{equation}
y=e^{i\delta} 
\end{equation}
would be scaled by this factor $\frac{2N_{\includegraphics[clip, width=1cm]{ooo.eps}}}{N_{\includegraphics[clip, width=1cm]{ooo.eps}}+N_{\includegraphics[clip, width=1cm]{wavy.line.eps}}}$ or
\begin{equation}
\delta\to \frac{2N_{\includegraphics[clip, width=1cm]{ooo.eps}}}{N_{\includegraphics[clip, width=1cm]{ooo.eps}}+N_{\includegraphics[clip, width=1cm]{wavy.line.eps}}}
\cdot \delta. 
\end{equation}
In this way we get the disc\\

\begin{figure}[htbp]
\begin{center}
\includegraphics[clip, width=5cm]{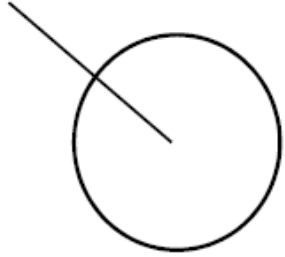} 
\end{center}
\caption{To get the wave functions of 
the strings $1$ and $3$, we make a functional integral
in a disc or half infinite cylinder Figure \ref{svms.34}
constructed described in the manner of the main text.}
\label{svms.40-upper}
\end{figure}

transformed rather into a ``hat'' or a cone\\

\begin{figure}[htbp]
\begin{center}
\includegraphics[clip, width=5cm]{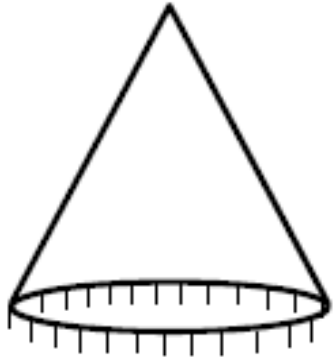} 
\end{center}
\caption{A half figure conformally transformed from
Figure \ref{svms.40-upper}}
We can scale up or down by changing the length of the edge with the objects attached to be.
\end{figure}

Now with the lengths of the edges with the object 
attachments reduced or scaled up to be rather than
of circle length $\pi$ of circle length
$\frac{2N_{\includegraphics[clip, width=1cm]{ooo.eps}}}{N_{\includegraphics[clip, width=1cm]{ooo.eps}}+N_{\includegraphics[clip, width=1cm]{wavy.line.eps}}}\cdot\pi$.
But after such scalings of the edges for the folded discs for 
both string $1$ and string $3$ we can -nicely- glue the $4$ previous discs, 
two of which are now hats, to actually topologically
speaking one Riemann sphere.\\
Since a disc and the complement of a disc are
conformally equivalent we can describe the result as the following ${\bf {\rm a), b), \ldots , f)}}$ of
the appropriate gluing of the regions for the
functional integral:\\
\begin{description}
\item[a)] Put the string $2$ disc as genuinely a unit
disc into the complex plane.
\item[b)] Take the string $4$ disc instead of as a genuine 
disc to be represented by the complement of a disc. Then the edges
for string $2$ and string $4$ lie just on the same unit circle.
But string $2$ and string $4$ have no pieces to be 
identified at all since $2$ only have 
\includegraphics[clip, width=1cm]{wavy.line.eps}
and \includegraphics[clip, width=1cm]{ooo.eps}
while $4$ has only \includegraphics[clip, width=1cm]{4.eps}
\includegraphics[clip, width=1cm]{xxx.eps}.
\item[c)] So we need to put the two hats produced with circular
curve lengths respectively
\begin{eqnarray}
{\rm 1-curvelength}=\frac{2N_{\includegraphics[clip, width=1cm]{ooo.eps}}}{N_{\includegraphics[clip, width=1cm]{ooo.eps}}+N_{\includegraphics[clip, width=1cm]{wavy.line.eps}}}\cdot\Pi\\ \nonumber
{\rm 3-curvelength}=\frac{2N_{\includegraphics[clip, width=1cm]{4.eps}}}{N_{\includegraphics[clip, width=1cm]{4.eps}}+N_{\includegraphics[clip, width=1cm]{xxx.eps}}}\cdot\Pi
\ {\rm in}.
\end{eqnarray}
Since according to our restrictions (during the summation)
\begin{equation}
N_{\includegraphics[clip, width=1cm]{ooo.eps}}=N_{\includegraphics[clip,width=1cm]{xxx.eps}} 
\end{equation}
and
\begin{equation}
N_{\includegraphics[clip, width=1cm]{wavy.line.eps}}=N_{\includegraphics[clip,width=1cm]{4.eps}} 
\end{equation}
we have to have ${\rm 1-curve length}+{\rm 3-curve length}=2\pi$
meaning that the curve lengths of the two ``hats'' just add up to $2\pi$,
the circumference unit circle seperating the disc for
string $2$ and the disc-complement for string $4$.
\item[d)] We shall attach the annihilation gluings.
That is to say that for example the hat for $1$ with its negative 
series \includegraphics[clip, width=1cm]{ooo.eps}
of objects to be attached to $2$.
\item[e)] Then the ``positive'' edge for $1$ named
 \includegraphics[clip, width=1cm]{xxx.eps}
can extends straight into the equally named 
 \includegraphics[clip, width=1cm]{xxx.eps}
for the disc complement for $4$.
\item[f)] Analogously the hat for $3$ should be
complement for $4$ along its ``negative''
 \includegraphics[clip, width=1cm]{4.eps}
piece. The positive part
 \includegraphics[clip, width=1cm]{wavy.line.eps}
 on the hat for $3$ then can go straight into the also
 \includegraphics[clip, width=1cm]{wavy.line.eps}
marked edge of the disc for $2$.
\end{description}
In the configuration just achieved the different 2-dimensional
pieced have be put so that the functional integral
variable $\phi^{\mu}$ should be analytic all over
in all the up to $3$ layers now being regions.\\
When we seek to construct a Riemann sphere or the like
we still have the problem that along the
internal separation lines in the hats $1$ and $3$,
You may reflect the layers of the ``negative'' type.\\

%

\section{Conclusion and outlook} 
We develop a new description for an arbitrary number of strings, a string field theory.\\
It is formulated in terms of a discretization into pieces - much like Thorn's 
string bits, but we do it for right and left movings- components and the then string bits of thorn are then called by us ``objects''.\\
These objects have dynmmics like free massless particles. That is to say they are decided by a quantum field theory of free massless particle. In momentum space they are static.\\
So nothing happens, even if the strings scatter!\\
We have arguments that our model  is really a transformation of theory for several strings.\\
As for deriving the Veneziano model, first with some troubles, but having negative even energy 
for the objects will presumably help to get the full Veneziano model (we missed two terms at first)\\
Also the spectrum we got o.k., except for a species doubler problem\cite{30}.\\

(Apart from null sets) our string field theory should be just a rewriting 
of usual say string field theory.\\
The Hilbert space describing all the possible states in a string world 
is the Fock space of -either one or two- theories of massless noninteracting 
scalars(for the bosonic $25+1$ model).\\
Two massless free scalar theories/species of scalar particles for purely closed string theory, 
while only one when there are open strings.\\
But allowed states are restricted to obey -approximately- some ``chiral'' 
invariant continuity condition: this means that the stringyness only 
comes in via initial state conditions.\\

We think we have a new(novel) way of representing string theory, 
which because of being in some respects simpler could be helpful 
in understanding some aspects of string theory better.\\
Even if string theory should not turn out to be the final truth 
-as can still be the case- its abilities for providing a cut off 
are so good that alone in looking for cut off it may give inspiration.\\
It happens generally thinking to seek a cut off you easily get 
in the direction of the string theory, especially the aspect of 
not having any true interaction as is a trademark for OUR NOVEL SFT MODEL.\\

Our novel field theory deviates from usual ones - Kaku Kikkawa's or 
Witten's by including (a nul set of) of information less in its 
description of state of the world, i.e. of a set of strings present.\\
We have rewritten the information - the kept part - on a state of 
several strings into a state of something (more like particles), 
which we call ``objects'', to such a degree that one only sees 
the connection to genuine strings by quite a bit of complicated rewriting.

Our novel string field theory is genuinely nonperturbative theory. We should be able to redenve nonperturbative theory of string theories such as branes. Also so fan the background space time is flat. Next step will be taking non flat, e.q. pp-wave background.

Now if our string field theory (and string theory) is the theory of everything (TOE), we should be able to derive inflation theories in early universe: it may be one of the greatest challenge which we are planning to attack. 

In very high energies such as Planck scale and/or string scale, we may be able to investigate truly new physics, for instance, studying supersymmetric particles.

\section*{Acknowledgements}
The authors thank K. Murakami, K. Sugiyama, M. Sakaguchi and
Y. Sekino for their useful comments.\\
One of us (H. B. N.) acknowledges the Niels Bohr
Institute for allowance to work as emeritus.  M. Ninomiya
acknowledges Yukawa Institute for Theoretical Physics, Kyoto University and Osaka city University, Advanced Mathematical Institute supporting this work. the Niels Bohr Institute and the Niels Bohr
International Academy for giving him very good hospitality 
during his stay.  M.N. also acknowldges the 
present research is supported in part by 
the JSPS Grant in Aid for Scientific Research 
No. 15K05063.\\
H.B.N. thanks to the Bled Conference participants,
organizers and Matiaz Breskov for finantial support to
come there where many of the ideas of this work got tested.

\appendix
\section{What is the Rough Dirac Sea?} 
In a free theory of second quantized fermions it is well known that 
the negative energy single fermion states are all filled while 
the positive energy single fermion states are all empty.\\
When there are interactions between such fermions or with other 
fields the ground state 
is no longer so simple. The vacuum is in this case 
rather a superposition of a lot of free energy states of the second quantized 
theory, a lot of which have empty single fermion negative energy 
states, or filled positive single fermion states.\\
This is analogous to that in a peaceful sea there is water for 
negative height and air for eigenstate positive height.\\
In a rough sea there is near height zero almost equal 
probability for finding water and air.\\
So if you act with an annihilation operator for a positive energy single
fermion state or with a negative single fermion creation operator 
on a vacuum with interaction, then you obtain a state, in which 
the sum of the single fermion energies (ignoring the interaction) has been lowered.\\
The interacting vacuum is by definition the lowest energy state, 
when the interaction is included, but it is not the lowest energy state 
for the free fermion energy, so the free fermion energy can easily be 
lowered by some annihilation of a positive energy fermion or 
creation of a negative energy one.\\
This is analogous to that you could remove a droplet of water from 
a positive height position from a rough sea; or you could add a droplet 
in a negative height place, with some slight amount of luck only needed.\\

\subsection{The Idea of the Rough Dirac Sea}
Really what we have in mind in the case of usual (particle)
quantum field theory under the notation of the ``Rough Dirac Sea''
is just the true vacuum of the quantum field theory in the case of a strongly
interacting theory.  In a free quantum field theory one has a Dirac sea in
which just all states with negative energy are filled while those with positive
energy are empty so that
\begin{eqnarray}
a(\vec{p},E>0)\mid 0 \rangle=0\\ \nonumber
a^{+}(\vec{p},E>0)\mid 0 \rangle\neq 0
\end{eqnarray}
and
\begin{eqnarray}
a_{anti}(\vec{p},E>0)\mid 0 \rangle&=&a^{+}(-\vec{p},-E<0)\mid 0 \rangle=0\\ \nonumber
a^{+}_{anti}(\vec{p},E>0)\mid 0 \rangle&=&a(-\vec{p},-E<0)\mid 0 \rangle\neq 0.
\end{eqnarray}
But now if there are interactions the true vacuum get much more complicated and 
one could obtain it by a development of some state, e.g. the bare vacuum $\mid 0 \rangle$
through a long imaginary time so that the propagation operator becomes
\begin{eqnarray}
e^{-H_{{t}_{large}}}
\end{eqnarray}
where $t_{large}$ is a very large time.  Then we get the true vacuum
\begin{equation}
\mid 0 \rangle_{true}\propto \lim_{t_{large}}\to \infty e^{-H_{{t}_{large}}}\mid 0 \rangle.
\end{equation}
If the interactions in Hamiltonian $H$ are strong the true vacuum is very much different from the bare 
one $\mid 0 \rangle$
Then it will be so that all operations with bare creation and annihilation operators
$a^{+
}(\vec{p},E \gtrless 0)$ and $a(\vec{p},E \gtrless 0)$ will give \underline{different} from zero
results when acting on $\mid 0 \rangle_{true}$.\\
I.e.
\begin{eqnarray}
a(\vec{p},E>0)\mid 0 \rangle_{true}&\neq& 0\\ \nonumber
a^{+}(\vec{p},E>0)\mid 0 \rangle_{true}&\neq& 0\\ \nonumber
a(\vec{p},E<0)\mid 0 \rangle_{true}&\neq& 0\\ \nonumber
a^{+}(\vec{p},E<0)\mid 0 \rangle_{true}&\neq& 0.
\end{eqnarray}
Thus it is quite easy seemingly to add to the true vacuum 
\underline{negative} energy. This is, however, only true when one
thinks of the free approximation energy $H_{0}$, then one has
\begin{equation}
\langle 0_{true}\mid a(\vec{p},E<0)H_{0}a^{+}(\vec{p},E<0)\mid 0_{true}\rangle < \langle 0 \mid_{true}H_{0}\mid 0 \rangle_{true}
\end{equation}
But using true Hamiltonian $H$ instead of $H_{0}$ would mean that the true vacuum is the
lowest eigenstate so that
\begin{equation}
\langle b \mid H \mid b \rangle \ge \langle 0 \mid_{ture} H \mid 0 \rangle_{true}
\end{equation}
for any state $\mid b \rangle$, also for 
\begin{equation}
\mid b \rangle=a^{+}(\vec{p},E<0)\mid 0 \rangle_{true}.
\end{equation}
Even if we decide to make the here considered fermion be a Majorana fermion
so that on the allowed states, the allowed subspace of the (second quantized)
Hilbert space we have
\begin{equation}
a^{+}_{anti}(\vec{p},E>0)\stackrel{=}{\scriptsize{\rm effectively}}a^{+}(\vec{p},E>0).
\end{equation}
We can still have this seeming - i.e. w.r.t. $H_{0}$ - addition of 
negative energy.\\
In our present work we want to use the story of the ``Rough Dirac sea''
for the objects ($\sim$ bits of strings taken in the $X_{R}$ and $X_{L}$
instead of Thorns $full X$) - which in our scheme are essentially
particles.\\
In our model we have, however, no interaction between objects and so
a genuine true vacuum is not obvious to define.  We so to speak lack
$H$ just above. However, we can without really explaining it assume that
there exist some especially selected - by God - background state
which we can identify with the true vacuum for the objects
$\mid 0 \rangle_{true\ (for\ obj)}$ and then we should think of
replacing the at first presented $\mid 0 >$ as vacuum for the objects by
 the more complicated $\mid 0 \rangle_{true\ (for\ obj)}$  
\begin{eqnarray}
\mid 0 \rangle \longrightarrow \mid 0 \rangle_{\begin{array}{c}{\scriptsize true {\rm (for\ obj)}} \end{array}}
\end{eqnarray}
Acting on this $\mid 0 \rangle_{true\ (for\ obj)}$ we can now seemingly
add both negative and positive energy. In this way one can on 
$\mid 0 \rangle_{true\ (for\ obj)}$ as background create states which
have negative energy density along the cyclic chains, and
they can be removed again by a positive energy preation
operator series: e.g.
\begin{equation}
\prod_{I}a^{+}(J^{\mu}(I))\mid 0 \rangle_{\begin{array}{c}{\scriptsize true {\rm(for\ obj)}}\end{array}}
\end{equation}
has compared $\mid 0 \rangle_{true\ (for\ obj)}$ itself
a negative ``bare'' energy $\Sigma_{J^{0}}$ if the $J^{\mu}$'s obey
\begin{equation}
J^{0}(I)<0.
\end{equation}

\section{Use of Rough Dirac sea Analogy for Our Novel String Field Theory Background State} 
Although we do not have any genuine interaction between the objects 
in our model, we shall nevertheless imagine that the ``background state'' 
on which we act with object-creation and object-annihilation operators 
is a complicated state, so that it is not the ground state for the sum 
of the single object states, so that it is not a problem to act with some 
annihilation or creation operator so as to add negative free energy, 
meaning make the sum of the object energies more negative.

%
%
%
%
%
%
%
%

%
%
%
%
%
%
%
\end{document}